\def\year{2021}\relax
\documentclass[letterpaper]{article} % DO NOT CHANGE THIS
\usepackage{aaai21}  % DO NOT CHANGE THIS
\usepackage{times}  % DO NOT CHANGE THIS
\usepackage{helvet} % DO NOT CHANGE THIS
\usepackage{courier}  % DO NOT CHANGE THIS
\usepackage[hyphens]{url}  % DO NOT CHANGE THIS
\usepackage{graphicx} % DO NOT CHANGE THIS
\usepackage{natbib}  % DO NOT CHANGE THIS AND DO NOT ADD ANY OPTIONS TO IT
\usepackage{caption} % DO NOT CHANGE THIS AND DO NOT ADD ANY OPTIONS TO IT
\usepackage{amssymb}
\usepackage{enumitem}
\usepackage{listings}
\usepackage{xstring}
\usepackage{graphicx}
\usepackage{pbox}
\usepackage{subcaption}
\usepackage{epstopdf}
\usepackage{xstring}

\usepackage{balance}
\usepackage{pbox}
\usepackage{multirow}
\usepackage{enumitem}

\usepackage{multicol}

\usepackage{tabularx}
\usepackage{tcolorbox}
\usepackage{color, colortbl}
\usepackage[show]{inotes}
\usepackage{soul}
\usepackage{booktabs}
\usepackage{multirow}
\usepackage[switch]{lineno}
\usepackage{amssymb}% http://ctan.org/pkg/amssymb
\usepackage{pifont}% http://ctan.org/pkg/pifont
\newcommand{\cmark}{\ding{51}}%
\newcommand{\xmark}{\ding{55}}%

\usepackage{comment}

\newcommand{\deltext}[1]{} %\sout{#1}}

\title{CrisisBench: Benchmarking Crisis-related Social Media Datasets\\ for Humanitarian Information Processing}
\author{Firoj Alam, Hassan Sajjad, Muhammad Imran, Ferda Ofli\\}
\affiliations{
    Qatar Computing Research Institute, HBKU, Qatar\\
    \{fialam,hsajjad,mimran,fofli\}@hbku.edu.qa
}

\begin{document}
% \linenumbers
\maketitle

\begin{abstract}
Time-critical analysis of social media streams is important for humanitarian organizations for planing rapid response during disasters. The \textit{crisis informatics} research community has developed several techniques and systems for processing and classifying big crisis-related data posted on social media. However, due to the dispersed nature of the datasets used in the literature (e.g., for training models), it is not possible to compare the results and measure the progress made towards building better models for crisis informatics tasks. In this work, we attempt to bridge this gap by combining various existing crisis-related datasets. We consolidate eight human-annotated datasets and provide 166.1k and 141.5k tweets for \textit{informativeness} and \textit{humanitarian} classification tasks, respectively. We believe that the consolidated dataset will help train more sophisticated models. Moreover, we provide benchmarks for both binary and multiclass classification tasks using several deep learning architecrures including, CNN, fastText, and transformers. %To that end, we provide binary and multiclass classification results using CNN, fastText, and transformer based models to address informativeness and humanitarian tasks, respectively. 
We make the dataset and scripts available at: \textit{\url{https://crisisnlp.qcri.org/crisis_datasets_benchmarks.html}}
\end{abstract}

\section{Introduction}
\label{sec:introduction}
At the onset of a disaster event, information pertinent to situational awareness such as reports of injured, trapped, or deceased people, urgent needs of victims, and infrastructure damage reports are most needed by formal humanitarian organizations to plan and launch relief operations. Acquiring such information in real-time is ideal to understand the situation as it unfolds. However, it is challenging as traditional methods such as field assessments and surveys are time-consuming. Microblogging platforms such as Twitter have been widely used to disseminate situational and actionable information by the affected population. Although social media sources are useful in this time-critical setting, it is, however, challenging to parse and extract actionable information from big crisis data available on social media~\cite{castillo2016big}. 

The past couple of years have witnessed a surge in the research works that focus on analyzing the usefulness of social media data and developing computational models to extract actionable information. Among others, proposed computational techniques include, information classification, information extraction, and summarization~\cite{imran2015processing}. 
%rudra2018identifying
Most of these studies use one of the publicly available datasets~\cite{olteanu2014crisislex,imran2016lrec,alam2018crisismmd,Alam9381294,alam-etal-2018-domain,alam2021social} and either propose a new model or report higher performance of an existing model. Typical classification tasks in the community include (i) \textit{informativeness} (i.e., informative vs.\ not-informative messages), (ii) \textit{humanitarian information types} (e.g., affected individual reports, infrastructure damage reports), and (iii) \textit{event types} (e.g., flood, earthquake, fire). 

% \inote{R1: "Another challenge [...] is to tackle the duplicate content", but the reason why this should be addressed is only explained in section 3.2.} % mentioned in below

Despite the recent focus of the \textit{crisis informatics}\footnote{\url{https://en.wikipedia.org/wiki/Disaster_informatics}} research community to develop novel and more robust computational algorithms and techniques to process social media data, we observe several limitations in the current literature. \textit{First}, few efforts have been invested to develop standard datasets (specifically, train/dev/test splits) and benchmarks for the community to compare their results, models, and techniques. \textit{Secondly}, most of the published datasets are noisy, e.g., CrisisLex \cite{olteanu2014crisislex} contains duplicate and near-duplicate content, which produces misleading classification performance. Moreover, some datasets (e.g., CrisisLex) consist of tweets from several languages without any explicit language tag, to separate the data of a particular language of interest.

To address such limitations, in this paper, we aim to develop a standard social media dataset for disaster response that facilitates comparison between different modeling approaches and encourages the community to streamline their efforts towards a common goal. We consolidate eight publicly available datasets (see Section \ref{sec:dataset}). The resulting dataset is larger in size, has better class distribution compared to the individual datasets, and enables building of robust models that performs better for various tasks (i.e., informativeness and humanitarian) and datasets. 
%, which are all important data features for building better models.
%%%%%%% dataset and class-label issue- common class label set

The consolidation of datasets from different sources involves various standardization challenges. 
One of the challenges is the inconsistent class labels across various data sources. 
We map the class labels using their semantic meaning---a step performed by domain experts manually. Another challenge is to tackle the duplicate content that is present within or across datasets. There are three types of duplicates: {\em(i)} tweet-id based duplicates (i.e., same tweet appears in different datasets), {\em(ii)} content-based duplicates (i.e., tweets with different ids have same content), which usually happens when users copy-paste tweets, and {\em(iii)} near-duplicate content (i.e., tweets with similar content), which happens due to retweets or partial copy of tweets from other users. 
%The duplicate problem even appears within a dataset while combining data from different disaster events. For example, the same tweet collected during hurricane Harvey and Irma, and it appears because of the keyword `hurricane', and also because the events happened in an overlapping time period. 
We use cosine similarity between tweets to filter out various types of duplicates. 
%Moreover, we provide a language tag with each tweet using 
%To add language tags we used 
% Google API. 
In summary, the contributions of this study are as follows:
% \vspace{-0.4em}
\begin{itemize}[leftmargin=*]
    \item  We consolidate eight publicly available disaster-related datasets by manually mapping semantically similar class labels, which leads to a larger dataset. 
    \item We carefully cleaned various forms of duplicates, and assigned a language tag to each tweet. 
    \item We provide benchmark results \textcolor{black}{on English tweets set} using state-of-the-art machine learning algorithms such as Convolutional Neural Networks (CNN), fastText~\citep{joulin2017bag} and pre-trained transformer models~\cite{devlin2018bert} for two classifications tasks, i.e., \textit{Informativeness} (binary) and \textit{Humanitarian type} (multi-class) classification.\footnote{\textcolor{black}{We only focused on two tasks for this study and we aim to address \textit{event types} task in a future study. 
    }} The benchmarking encourages the community towards a comparable and reproducible research.
    \item For the research community, we aim to release the dataset in multiple forms as, {\em(i)} a consolidated class label mapped version, {\em(ii)} exact- and near-duplicate filtered version obtained from previous versions, {\em(iii)} a subset of the filtered data used for the classification experiments in this study. 
    %Our released dataset includes a language tag, which enables the use of multilingual information in classification and is a promising future research direction.
    %conduct multilingual experiments as in future research.
\end{itemize}

% Available datasets and the lack of benchmarks...

% Contributions...
% - data consolidation
%     - merging of semantically similar classes
%     - content-based deduplication across datasets
%     - Reverse label propagation
% - benchmarking
%     - CNN
%     - BERT
%     - BERT with event info
%     - Multitask

%%%%%%%paper structure
The rest of the paper is organized as follows. Section~\ref{sec:related_works} provides an overview of the existing work. Section~\ref{sec:dataset} describes our data curation and consolidation procedures, and Section~\ref{sec:experiment_types} describes the experiments. Section~\ref{sec:results_discussion_future_works} presents and discusses the results. Finally, Section~\ref{sec:conclutions} concludes the paper.

\section{Related Work}
\label{sec:related_works}
% \vspace{-0.3em}

\subsection{Dataset Consolidation:}
% As discussed earlier most of the studies use one of the publicly available datasets, however, the datasets are limited for deep learning algorithms due to their size. In this study, we explore how all publicly available datasets can be combined, filtered and cleaned of exact- or near-duplicate content. Our data consolidation approach is similar to the studies of \cite{Alam2019} and \cite{kersten2019}. However, the study of \cite{Alam2019} is limited with respect to the fact that it has not been considered the issue of duplicate and near-duplicate content while combined different datasets. Whereas the study of \cite{kersten2019} only focused on informativeness classification\footnote{Note that in the literature, informativeness classification is also referred to as related vs.\ not-related.}.
% One of the main contributions of this study is the consolidating the publicly available datasets while filtering the exact- or near-duplicate content. 

In \textit{crisis informatics} research on social media, there has been an effort to develop datasets for the research community. An extensive literature review can be found in~\cite{imran2015processing}. Although there are several publicly available datasets that are used by the researchers, their results are not exactly comparable due to the differences in class labels and train/dev/test splits. In addition, the issue of exact- and near-duplicate content in existing datasets can lead to misleading performance as mentioned earlier. This problem become more visible while consolidating existing datasets. \citet{Alam2019,kersten2019} and \citet{Wiegmann2020} have previously worked in the direction to consolidate social media disaster response data. %However, both of these studies have limitations because 
A major limitation of the work by \citet{Alam2019} is that  the issue of duplicate and near-duplicate content have not been addressed when combining the different datasets. This issue resulted in an overlap between train and test sets. In terms of label mapping the work of \citet{Alam2019} is similar to the current study. \citet{kersten2019} focused only on informativeness\footnote{Authors used \textit{related} vs.\ \textit{not-related} in their study.} classification and combined five different datasets. This study has also not focused on exact- and near-duplicate content, which exist in different datasets. The study in \citet{Wiegmann2020} also compiled existing resources for \textit{disaster event types} (e.g., Flood, Fire) classification, which consists of a total of 123,166 tweets from 46 disasters with 9 disaster types. This is different from our work as we address \textit{informativeness} and \textit{humanitarian} classification tasks. Addressing \textit{disaster event types} classification is beyond the scope of our current study.
    
A fair comparison of the classification experiment is also difficult with previous studies as their train/dev/test splits are not public, except the dataset by \citet{Wiegmann2020}. We address such limitations in this study, i.e., we consolidate the datasets, eliminate duplicates, and release standard dataset splits with benchmark results.

In terms of defining class labels (i.e., tagsets)
%\footnote{We provided the list of class labels that are annotated in different datasets in the supplemental documents, Section \ref{ssec:mapped-class-label}.} 
for crisis informatics, most of the earlier efforts are discussed in~\cite{imran2015processing,temnikova2015emterms,stowe2018developing,Wiegmann2020}. Various recent studies~\cite{olteanu2014crisislex,imran2016lrec,alam2018crisismmd,stowe2018developing} use similar definitions for class labels. Unlike them,~\cite{strassel2017situational} defines more fine-grained categories based on need types (e.g., evacuation, food supply) and issue type (e.g., civil unrest). In this study, we use the class labels that are important for humanitarian aid for disaster response tasks, which are common across the publicly available resources. Some of the real-time applications that are currently using such labels include AIDR \cite{imran2014aidr}, CREES \cite{burel2018crisis}, and TweetTracker \cite{kumar2011tweettracker}.

\subsection{Classification Algorithms:}
Despite the fact that a majority of studies in \textit{crisis informatics} literature employ traditional machine learning algorithms, several recent works explore deep learning algorithms in disaster-related tweet classification tasks. The study of \cite{nguyen2017robust} and \cite{neppallideep} performed comparative experiments between different classical and deep learning algorithms including Support Vector Machines, Logistic Regression, Random Forests, Recurrent Neural Networks, and Convolutional Neural Networks (CNN). Their experimental results suggest that CNN outperforms other algorithms. Though in another study, \cite{burel2018crisis} reports that SVM and CNN can provide very competitive results in some cases. 
CNNs have also been explored in event type-specific filtering model~\cite{kersten2019} and few-shot learning~\cite{kruspedetecting2019}. Very recently different types of embedding representations have been proposed in literature such as Embeddings from Language Models (ELMo)~\cite{peters2018deep}, Bidirectional Encoder Representations from Transformers (BERT)~\cite{devlin2018bert}, and XLNet~\cite{yang2019xlnet} for different NLP tasks. The study by \cite{jain2019estimating} and \cite{Wiegmann2020} investigates these embedding representations for disaster tweet classification tasks.

\section{Data Curation}
\label{sec:dataset}

\subsection{Data Consolidation}
\label{ssec:datasets}
We consolidate eight datasets that were labeled for different disaster response classification tasks and whose labels can be mapped consistently for two tasks: \textit{informativeness} and \textit{humanitarian information type} classification. In doing so, we deal with two major challenges: {\em(i)} discrepancies in the class labels used across different datasets, and {\em(ii)} exact- and near-duplicate content that exists within as well as across different datasets.  
% We include CrisisLex (CrisisLexT6~\cite{olteanu2014crisislex}, CrisisLexT26~\cite{olteanu2015expect}), CrisisNLP~\cite{imran2016lrec}, SWDM2013~\cite{imran2013practical}, ISCRAM13 \cite{imran2013extracting}, Disaster Response Data (DRD),\footnote{\url{https://appen.com/datasets/combined-disaster-response-data/}} Disasters on Social Media (DSM),\footnote{\url{https://data.world/crowdflower/disasters-on-social-media}} CrisisMMD~\cite{alam2018crisismmd}, and data collected by AIDR system~\cite{imran2014aidr}.\footnote{Note that the AIDR system data has been annotated by domain experts and is available upon request.} 
Below we provide a brief overview of the datasets we used for consolidation. 
\begin{enumerate}
    \item \textbf{CrisisLex} is one of the largest publicly-available datasets, which consists of two subsets, i.e., CrisisLexT26 and CrisisLexT6~\cite{olteanu2014crisislex}. CrisisLexT26 comprises data from 26 different crisis events that took place in 2012 and 2013 with annotations for informative vs.\ not-informative as well as humanitarian categories (six classes) classification tasks among others. CrisisLexT6, on the other hand, contains data from six crisis events that occurred between October 2012 and July 2013 with annotations for \textit{related} vs.\ \textit{not-related} binary classification task.
    \item \textbf{CrisisNLP} is another large-scale dataset collected during 19 different disaster events that happened between 2013 and 2015, and annotated according to different schemes including classes from humanitarian disaster response and some classes related to health emergencies~\cite{imran2016lrec}.
    \item \textbf{SWDM2013} dataset consists of data from two events: {\em(i)} the Joplin collection contains tweets from the tornado that struck Joplin, Missouri on May 22, 2011; {\em(ii)} The Sandy collection contains tweets collected from Hurricane Sandy that hit Northeastern US on Oct 29, 2012~\cite{imran2013practical}.
    \item \textbf{ISCRAM2013} dataset consists of tweets from two different events occurred in 2011 (Joplin 2011) and 2012 (Sandy 2012). Note that this set of tweets are different than SWDM2013 set even though they are collected from same events \cite{imran2013extracting}. 
    %The Joplin 2011 data consists of 4,400 labeled tweets collected during the tornado that struck Joplin, Missouri (USA) on May 22, 2011, whereas Sandy 2012 data consists of 2,000 labeled tweets collected during the Hurricane Sandy, that hit Northeastern US on Oct 29, 2012.
    \item \textbf{Disaster Response Data (DRD)} consists of tweets collected during various crisis events that took place in 2010 and 2012. This dataset is annotated using 36 classes that include informativeness as well as humanitarian categories.\footnote{\url{https://appen.com/datasets/combined-disaster-response-data/}}
    \item \textbf{Disasters on Social Media (DSM)} dataset comprises 10K tweets collected and annotated with labels \textit{related} vs.\ \textit{not-related} to the disasters.\footnote{\url{https://data.world/crowdflower/disasters-on-social-media}}
    \item \textbf{CrisisMMD} is a multimodal dataset consisting of tweets and associated images collected during seven disaster events that happened in 2017~\cite{alam2018crisismmd}. The annotations for this dataset is targeted for three classification tasks: {\em (i)} informative \textit{vs.} not-informative, {\em (ii)} humanitarian categories (eight classes) and {\em (iii)} damage severity assessment.
    \item \textbf{AIDR} dataset is obtained from the \textit{AIDR system}~\cite{imran2014aidr} that has been annotated by domain experts for different events and made available upon requests. We only retained labeled data that are relevant to this study.
\end{enumerate}

First part of Table~\ref{table:data_sources_with_filtering} summarizes original sizes of the datasets. The CrisisLex and CrisisNLP datasets are the largest and second-largest datasets, respectively, which are currently widely used in the literature. The SWDM2013 is the smallest set. However, it is one of the earliest datasets used by the crisis informatics community.

\begin{table}[]
\centering
\scalebox{0.80}{
% \resizebox{\columnwidth}{!}{%
\begin{tabular}{@{}lr|rr|rr@{}}
\toprule
\multicolumn{1}{@{}l}{\textbf{Source}} & \multicolumn{1}{c|}{\textbf{Total}} & \multicolumn{2}{c|}{\textbf{Mapping}} & \multicolumn{2}{c}{\textbf{Filtering}} \\ \midrule
\multicolumn{1}{c}{\textbf{}} & \multicolumn{1}{c}{\textbf{}} & \multicolumn{1}{|c}{\textbf{Info}} & \multicolumn{1}{c}{\textbf{Hum}} & \multicolumn{1}{|c}{\textbf{Info}} & \multicolumn{1}{c@{}}{\textbf{Hum}} \\\midrule
CrisisLex & 88,015 & 84,407 & 84,407 & 69,699 & 69,699 \\
CrisisNLP & 52,656 & 51,271 & 50,824 & 40,401 & 40,074 \\
SWDM13 & 1,543 & 1,344 & 802 & 857 & 699 \\
ISCRAM13 & 3,617 & 3,196 & 1,702 & 2,521 & 1,506 \\
DRD & 26,235 & 21,519 & 7,505 & 20,896 & 7,419 \\
DSM & 10,876 & 10,800 & 0 & 8,835 & 0 \\
CrisisMMD & 16,058 & 16,058 & 16,058 & 16,020 & 16,020 \\
AIDR & 7,411 & 7,396 & 6,580 & 6,869 & 6,116 \\ \midrule
\textbf{Total} & \textbf{206,411} & \textbf{195,991} & \textbf{167,878} & \textbf{166,098} & \textbf{141,533} \\ \bottomrule
\end{tabular} 
}
% \vspace{-0.2em}
\caption{Different datasets and their sizes (number of tweets) before and after label mapping and filtering steps. Info: Informativeness, Hum: Humanitarian}
% \vspace{-0.5cm}
\label{table:data_sources_with_filtering}
\end{table}

\subsection{Class Label Mapping}
\label{ssec:class_label_mapping}
% \vspace{-0.4em}
The datasets come with different class labels. We create a set of common class labels by manually mapping semantically similar labels into one cluster.  For example, the label ``building damaged,'' originally used in the AIDR system, is mapped to ``infrastructure and utilities damage'' in our final dataset. Some of the class labels in these datasets are not annotated for \textit{humanitarian aid}\footnote{\url{https://en.wikipedia.org/wiki/Humanitarian_aid}} purposes, therefore, we have not included them in the consolidated dataset. For example, we do not select tweets labeled as ``animal management'' or ``not labeled'' that appear in CrisisNLP and CrisisLex26.
This causes a drop in the number of tweets for both informativeness and humanitarian tasks as can be seen in Table~\ref{table:data_sources_with_filtering} (Mapping column). The large drop in the CrisisLex dataset for the informativeness task is due to the 3,103 unlabeled tweets (i.e., labeled as ``not labeled''). The other significant drop for the informativeness task is in the DRD dataset. This is because many tweets were annotated with multiple labels, which we have not included in our consolidated dataset. \textcolor{black}{The reason is to reduce additional manual effort as it requires relabeling them for multiclass setting.}
Moreover, many tweets in these datasets were labeled for informativeness only. For example, the DSM dataset is only labeled for informativeness, and a large portion of the DRD dataset is labeled for informativeness only. We could not map them for the humanitarian task. 
%More details of this mapping can be found in the supplemental document \textcolor{red}{(Section \ref{ssec:mapped-class-label})}. 

\subsection{Exact- and Near-Duplicate Filtering}
\label{ssec:filtering}
To develop a machine learning model, it is important to design non-overlapping train/dev/test splits. A common practice is to randomly split the dataset into train/dev/test sets. This approach does not work with social media data as it generally contains duplicates and near duplicates. Such duplicate content, if present in both train and test sets, often leads to overestimated test results during classification. Filtering the near-and-exact duplicate content is one of the major steps we have taken into consideration while consolidating the datasets.

We first tokenize the text before applying any filtering. For tokenization, we used a modified version of the Tweet NLP tokenizer \cite{o2010tweetmotif}.\footnote{\url{https://github.com/brendano/ark-tweet-nlp}} Our modification includes lowercasing the text and removing URL, punctuation, and user id mentioned in the text. We then filter tweets having only one token.
Next, we apply exact string matching to remove exact duplicates. 
An example of an exact duplicate tweet is: ``\textit{RT \@Reuters: BREAKING NEWS: 6.3 magnitude earthquake strikes northwest of Bologna, Italy: USGS}'', which appear three times with exact match in CrisisLex26~\cite{olteanu2014crisislex} dataset that has been collected during Northern Italy Earthquakes, 2012.\footnote{\url{http://en.wikipedia.org/wiki/2012_Northern_Italy_earthquakes}}

\begin{figure}[t]
\centering
\includegraphics[width=0.9\linewidth]{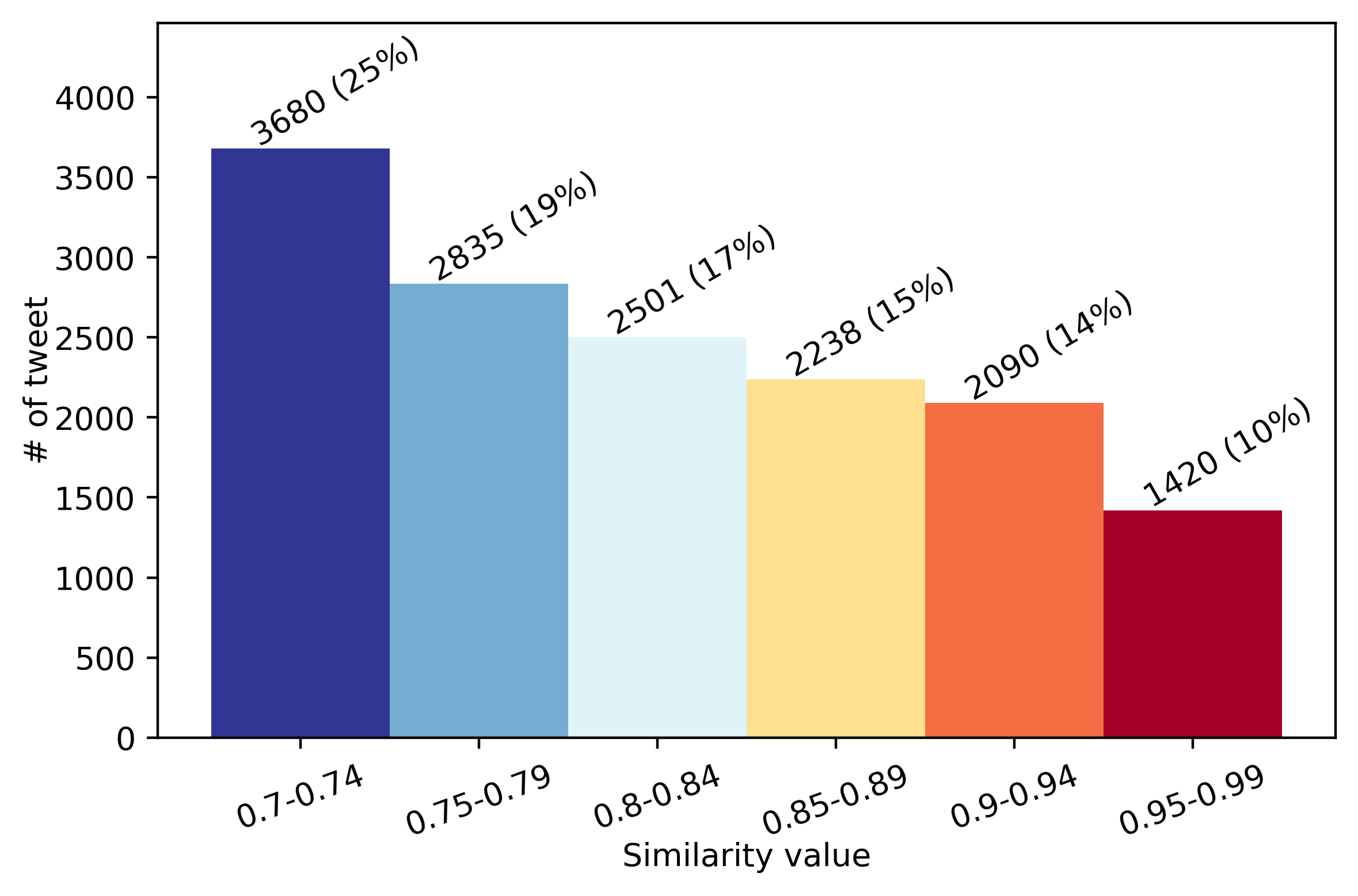}
% \vspace{-0.9em}
\caption{Number of near-duplicates in different bins obtained from consolidated informativeness tweets after label mapping. Tweets will lower similarity ($<0.7$) bins are not reported here.}
\label{fig:informativeness_sim_duplicates}
% \vspace{-1.0em}
\end{figure}

Then, we use a similarity-based approach to remove the near-duplicates. To do this, we first convert the tweets into vectors using bag-of-ngram approach as a vector representation. We use uni- and bi-grams with their frequency-based representations. We then use cosine similarity to compute a similarity score between two tweets and flag them as \emph{duplicate} (first three tweets in Table \ref{table:near-duplicates-examples}) if their similarity score is greater than the threshold value of 0.75.
In the similarity-based approach, threshold selection is an important aspect. Choosing a lower value would remove many distant tweets while choosing a higher value would leave several near-duplicate tweets in the dataset. \textcolor{black}{To determine a plausible threshold value, we manually checked the tweets in different threshold bins (i.e., 0.70 to 1.0 with 0.05 interval) as shown in Figure~\ref{fig:informativeness_sim_duplicates}, which we selected from consolidated informativeness dataset. By investigating the distribution and manual checking, we concluded that a threshold value of 0.75 is a reasonable choice. From the figure we can clearly see that choosing a lower threshold (e.g., $<0.75$) removes larger number of tweets. Note that rest of the tweets have similarity lower than what we have reported in the figure.} In Table~\ref{table:near-duplicates-examples}, we provide a few examples for the sake of clarity. 
%More examples can be found in the supplemental document (Section \ref{ssec:more_sim_examples}).

\begin{table*}[h!]
\centering
\scalebox{0.72}{
\begin{tabular}{lllll}
\toprule
\multicolumn{1}{l}{\textbf{\#}} & \multicolumn{1}{c}{\textbf{Tweet}} & \multicolumn{1}{c}{\textbf{Tokenized}} & \multicolumn{1}{c}{\textbf{Sim.}} & \multicolumn{1}{c}{\textbf{Dup.}}\\ \midrule

\multirow{2}{*}{1} & \parbox{8cm}{Live coverage: Queensland flood crisis via @Y7News http://t.co/Knb407Fw} & \parbox{8cm}{live coverage queensland flood crisis via url} & \multirow{2}{*}{0.788} & \multirow{2}{*}{\cmark}\\
 & \cellcolor[gray]{0.8}\parbox{8cm}{Live coverage: Queensland flood crisis - Yahoo!7 http://t.co/U2hw0LWW  via @Y7News} & \cellcolor[gray]{0.8}\parbox{8cm}{live coverage queensland flood crisis yahoo url via} &  \\ \midrule
 
\multirow{2}{*}{6} & \parbox{8cm}{Halo tetangga. Sabar ya. RT  @AJEnglish: Flood worsens in eastern Australia http://t.co/YfokqBmG} & \parbox{8cm}{halo tetangga sabar ya rt flood worsens in eastern australia url} & \multirow{2}{*}{0.787} & \multirow{2}{*}{\cmark}\\
 & \cellcolor[gray]{0.8}\parbox{8cm}{RT @AJEnglish: Flood worsens in eastern Australia http://t.co/kuGSMCiH} & \cellcolor[gray]{0.8}\parbox{8cm}{rt flood worsens in eastern australia url} &  \\ \midrule
\multirow{2}{*}{7} & \parbox{8cm}{"@guardian: Queensland counts flood cost as New South Wales braces for river peaks http://t.co/MpQskYt1". Brisbane friends moved to refuge.} & \parbox{8cm}{queensland counts flood cost as new south wales braces for river peaks url brisbane friends moved to refuge} & \multirow{2}{*}{0.778} & \multirow{2}{*}{\cmark}\\
 & \cellcolor[gray]{0.8}\parbox{8cm}{Queensland counts flood cost as New South Wales braces for river peaks http://t.co/qb5UuYf9} & \cellcolor[gray]{0.8}\parbox{8cm}{queensland counts flood cost as new south wales braces for river peaks url} & \\ \midrule
 
 \multirow{2}{*}{8} & \parbox{8cm}{RT @FoxNews: \#BREAKING: Numerous injuries reported in large explosion at \#Texas fertilizer plant http://t.co/oH93niFiAS". Brisbane friends moved to refuge.} & \parbox{8cm}{rt breaking numerous injuries reported in large explosion at texas fertilizer plant url} & \multirow{2}{*}{0.744} & \multirow{2}{*}{\xmark}\\ 
 & \cellcolor[gray]{0.8}\parbox{8cm}{Numerous injuries reported in large explosion at Texas fertilizer plant: DEVELOPING: Emergency crews in Texas ... http://t.co/Th5Yzvdg5m} & \cellcolor[gray]{0.8}\parbox{8cm}{numerous injuries reported in large explosion at texas fertilizer plant developing emergency crews in texas url} & \\ %   \midrule

 \bottomrule 
\end{tabular}
}
% \vspace{-0.7em}
\caption{Examples of near-duplicates with similarity scores selected from informativeness tweets. Duplicates are highlighted. \textit{Sim.} refers to similarity value. \textit{Dup.} refers to whether we consider them as duplicate. The symbol (\cmark) indicates a duplicate, which we dropped and the symbol (\xmark) indicates a non-duplicate, which we kept in our dataset.}
\label{table:near-duplicates-examples}
% \vspace{-1.0em}
\end{table*}

\begin{figure}[h!]
\centering
\includegraphics[width=0.9\linewidth]{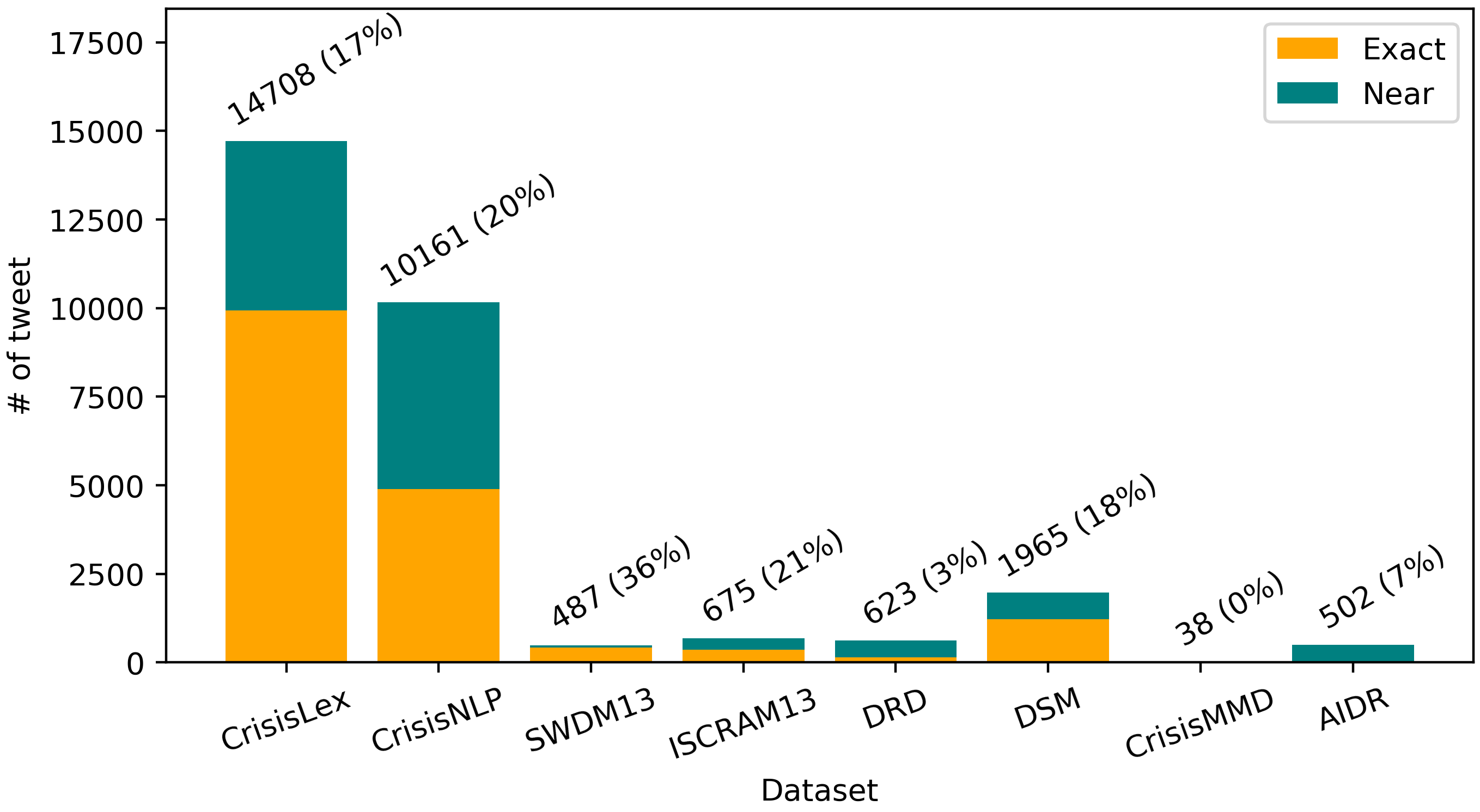}
% \vspace{-1.0em}
\caption{Exact- and near-duplicates in informativeness tweets. Number on top of each bar represents total number, and the number in the parenthesis represents percentage. %of consolidated exact and near duplicates from the respective dataset.
}
\label{fig:duplicates_exact_string_and_similarity}
% \vspace{-1.0em}
\end{figure}

We analyzed the data to understand which events and datasets have more exact- and near-duplicates. Figure \ref{fig:duplicates_exact_string_and_similarity} provides counts for both exact- and near-duplicates for informativeness tweets. In the figure, we report total number (in parenthesis the number represents percentage of reduction) of duplicates (i.e., exact and near) for each dataset. The CrisisLex and CrisisNLP have higher number of duplicates comparatively, however, it is because those two are relatively larger in size. For each of these datasets, we analyzed different events where duplicates appear most. In CrisisLex, the majority of the exact duplicates appear in ``Queensland floods (2013)''\footnote{Event name refers to the event during which data has been collected by the respective data authors (see Section \ref{ssec:datasets}).} consisting of 2270 exact duplicates. The second majority is ``West Texas explosion (2013)'' event, which consists of 1301 duplicates. Compared to CrisisLex, the exact duplicates are low in CrisisNLP, and the majority of such duplicates appear in the ``Philippines Typhoon Hagupit (2014)'' event with 1084 tweets. For the humanitarian tweets, we observed a similar trend. 
%characteristics of Figure \ref{fig:duplicates_exact_string_and_similarity}.

As indicated in Table~\ref{table:data_sources_with_filtering}, there is a drop after filtering, e.g., ${\sim}$25\% for informativeness and ${\sim}$20\% for humanitarian tasks. It is important to note that failing to eradicate duplicates from the consolidated dataset would potentially lead to misleading performance results in the classification experiments.

\subsection{Adding Language Tags}
\label{ssec:language issue}
% \vspace{-0.4em}
Some of the existing datasets contain tweets in various languages (i.e., Spanish and Italian) without explicit assignment of a language tag.
% While combining the datasets, we realized that some of them contain tweets in different languages (i.e., Spanish and Italian) other than English. 
In addition, many tweets have code-switched (i.e., multilingual) content. For example, the following tweet has both English and Spanish text: \textit{``It's \#Saturday, \#treat yourself to our \#Pastel tres leches y compota de mora azul. https://t.co/WMpmu27P9X''}. Note that Twitter tagged this tweet as English whereas the Google language detector service tagged it as Spanish with a confidence score of $0.379$. 

\begin{figure}[h]
\centering
\includegraphics[width=0.9\linewidth]{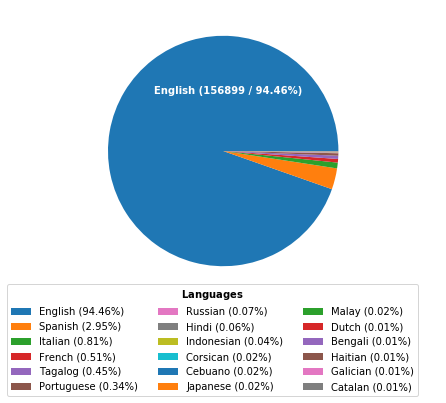}
% \vspace{-0.5em}
\caption{Distribution of top nineteen languages ($>=20$ tweets) in the consolidated informativeness tweets.}
\label{fig:lang_dist}
% \vspace{-0.6em}
\end{figure}

We decided to provide a language tag for each tweet if it is not available with the respective dataset. 
%For example, the tweets annotated by volunteers in the CrisisNLP dataset have language tags provided by Twitter whereas no language tag is provided with the CrisisLex dataset. For these tweets, 
We used the language detection API of Google Cloud Services\footnote{\url{https://cloud.google.com/translate/docs/advanced/detecting-language-v3}. Note, it is a paid service, therefore, we have not used this service for the tweets for which language tags are available.} for this purpose. 
%to assign the language tags. 
%We provided a language tag and confidence score obtained from the language detection API. 
%Hence, with the consolidated dataset, we include a language tag for all tweets. 
% In Figure~\ref{fig:lang_dist}, we report the distribution of languages with more than 20 tweets in the datasets. 
In Figure \ref{fig:lang_dist}, we report language distribution for the top nineteen languages consists of more than 20 tweets. Among different languages of informativeness tweets, English tweets appear to be highest in the distribution compared to any other language, which is 94.46\% of 156,899. Note that most of the non-English tweets appear in the CrisisLex dataset. We believe our language tags will enable future studies to perform multilingual analysis.

\begin{table*}[h]
\centering
\scalebox{0.76}{
\begin{tabular}{@{}lrrrrrrrrr@{}}
\toprule
\multicolumn{1}{@{}l}{\textbf{Class}} & \multicolumn{1}{r}{\textbf{CrisisLex}} & \multicolumn{1}{r}{\textbf{CrisisNLP}} & \multicolumn{1}{r}{\textbf{SWDM13}} & \multicolumn{1}{r}{\textbf{ISCRAM13}} & \multicolumn{1}{r}{\textbf{DRD}} & \multicolumn{1}{r}{\textbf{DSM}} & \multicolumn{1}{r}{\textbf{CrisisMMD}} & \multicolumn{1}{r}{\textbf{AIDR}} & \multicolumn{1}{r@{}}{\textbf{Total}} \\ \midrule
Informative & 42,140 & 23,694 & 716 & 2,443 & 14,849 & 3,461 & 11,488 & 2,968 & 101,759 \\
Not informative & 27,559 & 16,707 & 141 & 78 & 6,047 & 5,374 & 4,532 & 3,901 & 64,339 \\\midrule
\textbf{Total} & \textbf{69,699} & \textbf{40,401} & \textbf{857} & \textbf{2,521} & \textbf{20,896} & \textbf{8,835} & \textbf{16,020} & \textbf{6,869} & \textbf{166,098} \\ \bottomrule
\end{tabular}
}
% \vspace{-0.6em}
\caption{Data (\textit{\textbf{tweets containing multiple languages}}) distribution of \textit{\textbf{informativeness}} across different sources.}
%\vspace{-0.3cm}
\label{table:data_info_source_dist}
% \vspace{-0.7em}
\end{table*}

\begin{table*}[h]
\centering
\scalebox{0.76}{
\begin{tabular}{@{}lrrrrrrrr@{}}
\toprule
\multicolumn{1}{@{}l}{\textbf{Class}} & \multicolumn{1}{r}{\textbf{CrisisLex}} & \multicolumn{1}{r}{\textbf{CrisisNLP}} & \multicolumn{1}{r}{\textbf{SWDM13}} & \multicolumn{1}{r}{\textbf{ISCRAM13}} & \multicolumn{1}{r}{\textbf{DRD}} & \multicolumn{1}{r}{\textbf{CrisisMMD}} & \multicolumn{1}{r}{\textbf{AIDR}} & \multicolumn{1}{r@{}}{\textbf{Total}} \\ \midrule
Affected individual & 3,740 & - & - & - & - & 471 & - & 4,211 \\
Caution and advice & 1,774 & 1,137 & 117 & 412 & - & - & 161 & 3,601 \\
%\rowcolor[HTML]{DAE8FC}
\rowcolor[gray]{0.8}
Disease related & - & 1,478 & - & - & - & - & - & 1,478 \\
Displaced and evacuations & - & 495 & - & - & - & - & 50 & 545 \\
Donation and volunteering & 1,932 & 2,882 & 27 & 189 & 10 & 3,286 & 24 & 8,350 \\
Infrastructure and utilities damage & 1,353 & 1,721 & - & - & 877 & 1,262 & 283 & 5,496 \\
Injured or dead people & - & 2,151 & 139 & 125 & - & 486 & 267 & 3,168 \\
Missing and found people & - & 443 & - & 43 & - & 40 & 46 & 572 \\
Not humanitarian & 27,559 & 16,708 & 142 & 81 & - & 4,538 & 3,911 & 52,939 \\
\rowcolor[gray]{0.8}
Other relevant information & 29,562 & 8,188 & - & - & - & 5,937 & 939 & 44,626 \\
\rowcolor[gray]{0.8}
Personal update & - & 116 & 274 & 656 & - & - & - & 1,046 \\
\rowcolor[gray]{0.8}
Physical landslide & - & 538 & - & - & - & - & - & 538 \\
Requests or needs & - & 215 & - & - & 6,532 & - & 257 & 7,004 \\
Response efforts & - & 1,114 & - & - & - & - & - & 1,114 \\
Sympathy and support & 3,779 & 2,872 & - & - & - & - & 178 & 6,829 \\
\rowcolor[gray]{0.8}
Terrorism related & - & 16 & - & - & - & - & - & 16 \\ \midrule
\textbf{Total} & \multicolumn{1}{r}{\textbf{69,699}} & \multicolumn{1}{r}{\textbf{40,074}} & \multicolumn{1}{r}{\textbf{699}} & \multicolumn{1}{r}{\textbf{1,506}} & \multicolumn{1}{r}{\textbf{7,419}} & \multicolumn{1}{r}{\textbf{16,020}} & \multicolumn{1}{r}{\textbf{6,116}} & \multicolumn{1}{r@{}}{\textbf{141,533}} \\ \bottomrule
\end{tabular}
}
% \vspace{-0.5em}
\caption{Data (\textit{\textbf{tweets containing multiple languages}}) distribution of \textit{\textbf{humanitarian}} categories across different datasets.}
%\vspace{-0.3cm}
\label{table:data_hum_source_dist}
% \vspace{-1.0em}
\end{table*}

\subsection{Data Statistics}
\label{ssec:class_label_distribution}
% \vspace{-0.4em}
Distribution of class labels is an important factor for developing the classification model. In Table~\ref{table:data_info_source_dist} and \ref{table:data_hum_source_dist}, we report individual datasets along with the class label distribution for informativeness and humanitarian tasks, respectively. It is clear that there is an imbalance in class distributions in different datasets and some class labels are not present. For example, the distribution of ``not informative'' class is very low in SWDM2013 and ISCRAM2013 datasets. For the humanitarian task, some class labels are not present in different datasets. Only 17 tweets with the label ``terrrorism related'' are present in CrisisNLP. Similarly, the class ``disease related'' only appears in CrisisNLP. The scarcity of the class labels poses a great challenge to design the classification model using individual datasets. Even after combining the datasets, the imbalance in class distribution seems to persist (last column in Table \ref{table:data_hum_source_dist}). For example, the distribution of ``not humanitarian'' is relatively higher (37.40\%) than other class labels.
%, which might have to be under-sampled for training the classification model. 
In Table~\ref{table:data_hum_source_dist}, we highlighted some class labels, which we dropped in the rest of the classification experiments conducted in this study. However, tweets with those class labels will be available in the released datasets. The reason for not including them in the experiments is that we aim to develop classifiers for the disaster response tasks only.

% \begin{table}[!htp]
% \centering
% \scalebox{0.90}{
% \begin{tabular}{@{}lrrrr@{}}
% \toprule
% \textbf{Data Source} & \multicolumn{2}{l}{\textbf{Informativeness}} & \multicolumn{2}{l}{\textbf{Humanitarian}} \\ \midrule
%  & \multicolumn{1}{c}{Before} & \multicolumn{1}{c}{After} & \multicolumn{1}{c}{Before} & \multicolumn{1}{c}{After} \\\midrule
% CrisisLex & 82194 & 66954 & 43140 & 40627 \\
% CrisisNLP & 48942 & 38014 & 34759 & 31520 \\
% SWDM & 1344 & 881 & 506 & 445 \\
% DRD & 21507 & 20735 & 7503 & 7469 \\
% DSM & 10766 & 8673 & -- & -- \\
% CrisisMMD & 16058 & 15979 & 6384 & 6380 \\
% AIDR & 28170 & 7343 & 22319 & 7491 \\\midrule
% \textbf{Total} & \textbf{208981} & \textbf{158579} & \textbf{114611} & \textbf{93932} \\ \bottomrule
% \end{tabular}
% }
% \caption{Data source and their distribution before and after filtering}
% %\vspace{-0.3cm}
% \label{table:data_sources_with_filtering}
% \end{table}

%However, for the \textcolor{red}{multi-task classification} experiments, we use only CrisisMMD dataset as it provides aligned annotations for both informativeness and humanitarian tasks. \textcolor{red}{In addition, we do not remove tweets labeled as ``other relevant information'' for this experiment.}

\section{Experiments}
\label{sec:experiment_types}
% \vspace{-0.3em}

Although our consolidated dataset contains multilingual tweets, we only use tweets in English language in our experiments. We split data into train, dev, and test sets with a proportion of 70\%, 10\%, and 20\%, respectively, also reported in Table~\ref{tab:data_split_info_hum_tasks_combined_filtered}. As mentioned earlier we have not selected the tweets with highlighted class labels in Table~\ref{table:data_hum_source_dist} for the classification experiments. Therefore, in this and next Section \ref{sec:results_discussion_future_works} 
%the rest of the paper, 
we report the class label distribution and results on the selected class labels with English tweets only. 

\subsection{Models and Architectures}
\label{sec:experimental_design}
% \vspace{-0.4em}
In this section, we describe the details of our classification models. For the experiments, we use CNN, fastText, and pre-trained transformer models. 

\paragraph{CNN:}
The current state-of-the-art disaster classification model is based on the CNN architecture. We used similar architecture as proposed by \cite{nguyen2017robust}.

\paragraph{fastText:}
For the fastText~\citep{joulin2017bag}, we used pre-trained embeddings trained on Common Crawl, which is released by fastText for English. 
%\footnote{\url{https://commoncrawl.org/}}

\paragraph{Transformer models:}
Pre-trained models have achieved state-of-the-art performance on natural language processing tasks and have been adopted as feature extractors for solving down-stream tasks such as question answering, and sentiment analysis. Though the pre-trained models are mainly trained on non-Twitter text, we hypothesize that their rich contextualized embeddings would be beneficial for the disaster domain. In this work, we choose the pre-trained models BERT \cite{devlin2018bert}, DistilBERT \cite{sanh2019distilbert}, %XLM-RoBERTa~\citep{conneau2019unsupervised} 
and RoBERTa \citep{liu2019roberta} for the classification tasks.

\paragraph{Model Settings:}
\label{ssec:classificaiton_algo}
We train the CNN models using the Adam optimizer~\cite{kingma2014adam}. The batch size is 128 and maximum number of epochs is set to 1000. We use a filter size of 300 with both window size and pooling length of 2, 3, and 4, and a dropout rate 0.02. We set \emph{early stopping} criterion based on the accuracy of the development set with a patience of 200.
% \paragraph{FastText:} 
For the experiments with fastText, we used default parameters except: {\em(i)} the dimension is set to 300, {\em(ii)} minimal number of word occurrences is set to 3, and {\em(iii)} number of epochs is 50. 
% \paragraph{BERT:} 
We train transformers models using the Transformer Toolkit~\cite{Wolf2019HuggingFacesTS}. For each model, we use an additional task-specific layer. We fine-tune the model using fine-tuning procedure as prescribed by \cite{devlin2018bert}. Due to the instability of the pre-trained models as reported in~\cite{devlin2018bert}, we perform ten re-runs for each experiment using different seeds, and we select the model that performs best on the dev set. For transformer-based models, we used a learning rate of $2e-5$, and a batch size of 8. More details of the parameters setting can be found 
%in the supplemental document. (Section~\ref{sec:exp_parameters}) and 
in the released scripts.

\subsection{Preprocessing and Evaluation}
% \vspace{-0.2em}
\paragraph{Preprocessing:}
% \label{ssec:preprocessing_data}
Prior to the classification experiment, we preprocess tweets to remove symbols, emoticons, invisible and non-ASCII characters, punctuations (replaced with whitespace), numbers, URLs, and hashtag signs. 

\paragraph{Evaluation:}
To measure the performance of each classifier, we use weighted average precision (P), recall (R), and F1-measure (F1). The rationale behind choosing the weighted metric is that it takes into account the class imbalance problem. 

\subsection{Experimental Settings}
% \vspace{-0.2em}
\paragraph{Individual vs.\ Consolidated Datasets:}
\label{ssec:individual_dataset}
The motivation of these experiments is to investigate whether model trained with consolidated dataset generalizes well across different sets. For the individual dataset classification experiments, we selected CrisisLex and CrisisNLP as they are relatively larger in size and have a reasonable number of class labels, i.e., six and eleven class labels, respectively. Note that these are subsets of the consolidated dataset reported in Table~\ref{tab:data_split_info_hum_tasks_combined_filtered}. We selected them from train, dev and test splits of the consolidated dataset to be consistent across different classification experiments. To understand the effectiveness of the smaller datasets, we run experiments by training the model using smaller datasets and evaluating using the consolidated test set.

\begin{table}[!t]
\centering
\setlength{\tabcolsep}{3pt}
\scalebox{0.76}{
%\resizebox{\columnwidth}{!}{%
\begin{tabular}{lrrrr}
\toprule
\textbf{Informativeness} & \multicolumn{1}{c}{\textbf{Train}} & \multicolumn{1}{c}{\textbf{Dev}} & \multicolumn{1}{c}{\textbf{Test}} & \multicolumn{1}{c}{\textbf{Total}} \\ \midrule
Informative & 65826 & 9594 & 18626 & 94046 \\
Not informative & 43970 & 6414 & 12469 & 62853 \\ \midrule
\textbf{Total} & \textbf{109796} & \textbf{16008} & \textbf{31095} & \textbf{156899} \\ \midrule 
\multicolumn{5}{l}{\textbf{Humanitarian}} \\ \midrule
Affected individual & 2454 & 367 & 693 & 3514 \\
Caution and advice & 2101 & 309 & 583 & 2993 \\
Displaced and evacuations & 359 & 53 & 99 & \textbf{511} \\
Donation and volunteering & 5184 & 763 & 1453 & 7400 \\
Infrastructure and utilities damage & 3541 & 511 & 1004 & 5056 \\
Injured or dead people & 1945 & 271 & 561 & 2777 \\
Missing and found people & 373 & 55 & 103 & \textbf{531} \\
Not humanitarian & 36109 & 5270 & 10256 & 51635 \\
Requests or needs & 4840 & 705 & 1372 & 6917 \\
Response efforts & 780 & 113 & 221 & \textbf{1114} \\
Sympathy and support & 3549 & 540 & 1020 & 5109 \\ \midrule
\textbf{Total} & \textbf{61235} & \textbf{8957} & \textbf{17365} & \textbf{87557}  \\\bottomrule
\end{tabular}
}
% \vspace{-0.5em}
\caption{Data split and their distributions with the consolidated \textit{\textbf{English}} tweets dataset.}
\label{tab:data_split_info_hum_tasks_combined_filtered}
% \vspace{-1.0em}
\end{table}

\paragraph{Event-aware Training}
\label{ssec:event-aware} 
% \vspace{-0.2em}
The availability of annotated data for a disaster event is usually scarce. One of the advantages of our compiled data is to have identical classes across several disaster events. This enables us to combine the annotated data from all previous disasters for the classification. Though this increases the size of the training data substantially, the classifier may result in sub-optimal performance due to the inclusion of heterogeneous data (i.e., a variety of disaster types and occurs in a different part of the world). 
\citet{sennrich-haddow-birch:2016:N16-1} proposed a tag-based strategy where they add a tag to machine translation training data to force a specific type of translation. The method has later been adopted to do domain adaptation and multilingual machine translation~\cite{chu:acl2017,sajjad-etal:iwslt17}. Motivated by it, we propose an event-aware training mechanism. Given a set of $m$ disaster event types $\mathbf{D} = \{d_1, d_2, ... , d_m\}$ where disaster event type $d_i$ includes earthquake, flood, fire, and hurricane. For a disaster event type $d_i$, $\mathbf{T_i}$ = $\{t_1, t_2, ... , t_n\}$ are the annotated tweets. We append a disaster event type as a token to each annotated tweet $t_i$. More concretely, say tweet $t_i$ consists of $k$ words $\{w_1, w_2, ... , w_k\}$. We append a disaster event type tag $d_i$ to each tweet so that $t_i$ would become $\{d_i, w_1, w_2, ... , w_k\}$. We repeat this step for all disaster event types present in our dataset. We concatenate the modified data of all disasters and use it for the classification.

The event-aware training requires the knowledge of the disaster event type at the time of the test. If we do not provide a disaster event type, the classification performance will be suboptimal due to a mismatch between train and test. 
To apply the model to an unknown disaster event type, we modify the training procedure. Instead of appending the disaster event type to all tweets of a disaster, we randomly append disaster event type \texttt{UNK} to 5\% of the tweets of every disaster. Note that \texttt{UNK} is now distributed across all disaster event types and is a good representation of an unknown event.

\section{Results and Discussions}
\label{sec:results_discussion_future_works}
% \vspace{-0.4em}

\begin{table}[t!]
\centering
\setlength{\tabcolsep}{3pt}
\scalebox{0.74}{
\begin{tabular}{llrrrr}
\toprule
\multicolumn{1}{c}{\textbf{Train}} & \multicolumn{1}{c}{\textbf{Test}} & \multicolumn{1}{c}{\textbf{Acc}} & \multicolumn{1}{c}{\textbf{P}} & \multicolumn{1}{c}{\textbf{R}} & \multicolumn{1}{c}{\textbf{F1}} \\ \midrule
\multicolumn{6}{c}{\textbf{Informativeness}} \\ \midrule
CrisisLex (2C) & 1. CrisisLex & 0.945 & 0.945 & 0.950 & 0.945 \\
 & 2. CrisisNLP & 0.689 & 0.688 & 0.690 & 0.689 \\
 & 3. \cellcolor[HTML]{DAE8FC}Consolidated & \cellcolor[HTML]{DAE8FC}0.801 & \cellcolor[HTML]{DAE8FC}0.807 & \cellcolor[HTML]{DAE8FC}0.800 & \cellcolor[HTML]{DAE8FC}0.803 \\ \midrule
CrisisNLP (2C) & 4. CrisisNLP & 0.832 & 0.832 & 0.830 & 0.832 \\
 & 5. CrisisLex & 0.712 & 0.803 & 0.710 & 0.705 \\
 & 6. \cellcolor[HTML]{DAE8FC}Consolidated & \cellcolor[HTML]{DAE8FC}0.725 & \cellcolor[HTML]{DAE8FC}0.768 & \cellcolor[HTML]{DAE8FC}0.730 & \cellcolor[HTML]{DAE8FC}0.727 \\ \midrule
Consolidated (2C) & 7. CrisisLex & 0.943 & 0.943 & 0.940 & 0.943 \\
 & 8. CrisisNLP & 0.829 & 0.828 & 0.830 & 0.828 \\
 & 9. \cellcolor[HTML]{DAE8FC}Consolidated & \cellcolor[HTML]{DAE8FC}0.867 & \cellcolor[HTML]{DAE8FC}0.866 & \cellcolor[HTML]{DAE8FC}0.870 & \cellcolor[HTML]{DAE8FC}0.866 \\
\midrule
\multicolumn{6}{c}{\textbf{Humanitarian}} \\\midrule
CrisisLex (6C) & 10. CrisisLex & 0.921 & 0.920 & 0.920 & 0.920 \\
 & 11. CrisisNLP & 0.554 & 0.546 & 0.550 & 0.544 \\
 & 12. \cellcolor[HTML]{DAE8FC}Consolidated & \cellcolor[HTML]{DAE8FC}0.694 & \cellcolor[HTML]{DAE8FC}0.601 & \cellcolor[HTML]{DAE8FC}0.690 & \cellcolor[HTML]{DAE8FC}0.633 \\\midrule
CrisisNLP (10C) & 13. CrisisNLP & 0.780 & 0.757 & 0.780 & 0.762 \\
 & 14. CrisisLex & 0.769 & 0.726 & 0.770 & 0.714 \\
 & 15. \cellcolor[HTML]{DAE8FC}Consolidated & \cellcolor[HTML]{DAE8FC}0.666 & \cellcolor[HTML]{DAE8FC}0.582 & \cellcolor[HTML]{DAE8FC}0.670 & \cellcolor[HTML]{DAE8FC}0.613 \\ \midrule
Consolidated (11C) & 16. CrisisLex & 0.908 & 0.916 & 0.910 & 0.912 \\
 & 17. CrisisNLP & 0.768 & 0.753 & 0.770 & 0.753 \\
 & 18. \cellcolor[HTML]{DAE8FC}Consolidated & \cellcolor[HTML]{DAE8FC}0.835 & \cellcolor[HTML]{DAE8FC}0.827 & \cellcolor[HTML]{DAE8FC}0.840 & \cellcolor[HTML]{DAE8FC}0.829 \\
\bottomrule
\end{tabular}
}
% \vspace{-0.2em}
\caption{Classification results using \textbf{CNN} for the individual and consolidated datasets. 6C, 10C, and 11C refer to six, ten and eleven class labels respectively. }
\label{tab:individual_combined_datasets_cnn}
% \vspace{-1.0em}
\end{table}

\subsection{Individual vs.\ Consolidated Datasets}
\label{ssec:res_ind_vs_con}
In Table~\ref{tab:individual_combined_datasets_cnn}, we report the classification results for individual vs.\ consolidated datasets for both informativeness and humanitarian tasks using the CNN model. As mentioned earlier, we selected CrisisLex and CrisisNLP to conduct experiments for the individual datasets. The model trained with individual dataset shows that performance is higher on the corresponding set but low on other sets. For example, for informativeness task, the model trained with CrisisLex performs better on CrisisLex but not on CrisisNLP and Consolidated sets. We see similar pattern for CrisisNLP. However, the model trained with Consolidated data shows similar performance as individual sets (i.e., CrisisLex and CrisisNLP) but higher on consolidated set. A comparison is shown in highlighted rows in the Table \ref{tab:individual_combined_datasets_cnn}. The model trained using the consolidated dataset achieves 0.866 (F1) for informativeness and 0.829 for humanitarian, which is better than the models trained using individual datasets. This proves that model with consolidated dataset generalizes well, obtaining similar performance on individual sets and higher on the consolidated set.

Between CrisisLex and CrisisNLP, the performance is higher on CrisisLex dataset for both informativeness and humanitarian tasks (1st \textit{vs.} 4th row in Table \ref{tab:individual_combined_datasets_cnn} for the informativeness, and 10th \textit{vs.} 13th row for the humanitarian task in the same table.). This might be due to the CrisisLex dataset being larger than the CrisisNLP dataset. The cross dataset (i.e., train on CrisisLex and evaluate on CrisisNLP) results shows that there is a drop in performance. For example, there is 14.3\% difference in F1 on CrisisNLP data using the CrisisLex model for the informativeness task. Similar behavior observed when evaluated the CrisisNLP model on the CrisisLex dataset. In the humanitarian task, for different datasets in Table~\ref{tab:individual_combined_datasets_cnn}, we have different number of class labels. We report the results of those classes only for which the model is able to classify. For example, the model trained using the CrisisLex data can classify tweets using one of the six class labels (see Table~\ref{table:data_hum_source_dist} for excluded labels with highlights). The experiments with smaller datasets for both informativeness and humanitarian tasks show the importance of designing a classifier using a larger dataset. Note that humanitarian task is a multi-class classification problem, which makes it a much more difficult task than the binary informativeness classification.

\textbf{Comparing Models:}
In Table \ref{tab:combined_datasets_fastText_transformers}, we report the results using CNN, fastText and transformer based models. We report weighted F1 for all models and tasks. %with the model trained on individual and consolidated sets (as shown in column Train) and tested on consolidated test set. 
The transformer based models achieve higher performance compared to the CNN and fastText. We used three transformer based models, which varies in the parameter sizes. %(see Section \ref{sec:exp_parameters} in the supplemental document). 
However, in terms of performance, they are quite similar.

\begin{table}[]
\centering
\setlength{\tabcolsep}{3pt}
\scalebox{0.74}{
\begin{tabular}{@{}llrrrrr@{}}
\toprule
\multicolumn{1}{c}{\textbf{Train}} & \multicolumn{1}{c}{\textbf{Test}} & \multicolumn{1}{c}{\textbf{CNN}} & \multicolumn{1}{c}{\textbf{FT}} & \multicolumn{1}{c}{\textbf{BERT}} & \multicolumn{1}{c}{\textbf{D-B}} & \multicolumn{1}{c}{\textbf{RT}} \\ \midrule
\multicolumn{7}{c}{\textbf{Informativeness}} \\ \midrule
 & 1. CrisisLex  & 0.945 & 0.940 & 0.949 & 0.949 & 0.938 \\
 & 2. CrisisNLP & 0.689 & 0.687 & 0.698 & 0.681 & 0.694 \\
\multirow{-3}{*}{CrisisLex (2C)} & 3. \cellcolor[HTML]{DAE8FC}Consolidated & \cellcolor[HTML]{DAE8FC}0.803 & \cellcolor[HTML]{DAE8FC}0.791 & \cellcolor[HTML]{DAE8FC}0.806 & \cellcolor[HTML]{DAE8FC}0.808 & \cellcolor[HTML]{DAE8FC}0.810 \\ \midrule
 & 4. CrisisNLP & 0.832 & 0.816 & 0.833 & 0.834 & 0.823 \\
 & 5. CrisisLex & 0.705 & 0.728 & 0.749 & 0.739 & 0.726 \\
\multirow{-3}{*}{CrisisNLP (2C)} & 6. \cellcolor[HTML]{DAE8FC}Consolidated & \cellcolor[HTML]{DAE8FC}0.727 & \cellcolor[HTML]{DAE8FC}0.733 & \cellcolor[HTML]{DAE8FC}0.753 & \cellcolor[HTML]{DAE8FC}0.755 & \cellcolor[HTML]{DAE8FC}0.744 \\\midrule
 & 7. CrisisLex & 0.943 & 0.917 & 0.940 & 0.938 & 0.946 \\
 & 8. CrisisNLP & 0.828 & 0.811 & 0.825 & 0.828 & 0.829 \\
\multirow{-3}{*}{Consolidated (2C)} & 9. \cellcolor[HTML]{DAE8FC}Consolidated & \cellcolor[HTML]{DAE8FC}0.866 & \cellcolor[HTML]{DAE8FC}0.844 & \cellcolor[HTML]{DAE8FC}0.872 & \cellcolor[HTML]{DAE8FC}0.870 & \cellcolor[HTML]{DAE8FC}0.883 \\ \midrule 
\multicolumn{7}{c}{\textbf{Humanitarian}} \\\midrule
 & 10. CrisisLex & 0.920 & 0.911 & 0.934 & 0.935 & 0.937 \\
 & 11. CrisisNLP & 0.544 & 0.549 & 0.615 & 0.628 & 0.632 \\
\multirow{-3}{*}{CrisisLex (6C)} & 12. \cellcolor[HTML]{DAE8FC}Consolidated & \cellcolor[HTML]{DAE8FC}0.633 & \cellcolor[HTML]{DAE8FC}0.605 & \cellcolor[HTML]{DAE8FC}0.766 & \cellcolor[HTML]{DAE8FC}0.770 & \cellcolor[HTML]{DAE8FC}0.784 \\ \midrule
 & 13. CrisisNLP & 0.762 & 0.759 & 0.791 & 0.788 & 0.789 \\
 & 14. CrisisLex & 0.714 & 0.719 & 0.842 & 0.845 & 0.850 \\
\multirow{-3}{*}{CrisisNLP (10C)} & 15. \cellcolor[HTML]{DAE8FC}Consolidated & \cellcolor[HTML]{DAE8FC}0.613 & \cellcolor[HTML]{DAE8FC}0.627 & \cellcolor[HTML]{DAE8FC}0.709 & \cellcolor[HTML]{DAE8FC}0.707 & \cellcolor[HTML]{DAE8FC}0.727 \\\midrule
 & 16. CrisisLex & 0.912 & 0.903 & 0.923 & 0.921 & 0.931 \\
 & 17. CrisisNLP & 0.753 & 0.760 & 0.786 & 0.787 & 0.784 \\
\multirow{-3}{*}{Consolidated (11C)} & 18. \cellcolor[HTML]{DAE8FC}Consolidated & \cellcolor[HTML]{DAE8FC}0.829 & \cellcolor[HTML]{DAE8FC}0.824 & \cellcolor[HTML]{DAE8FC}0.860 & \cellcolor[HTML]{DAE8FC}0.856 & \cellcolor[HTML]{DAE8FC}0.872 \\
\bottomrule
\end{tabular}
}
% \vspace{-0.2em}
\caption{Classification results (weighted-F1) using CNN, fastText (FT) and transformer based models. D-B: DistilBERT, RT: RoBERTa.}
\label{tab:combined_datasets_fastText_transformers}
% \vspace{-1.0em}
\end{table}

\textbf{Class-wise Results Analysis:}
% \label{sec:class-wise_results}
In Table~\ref{tab:combined_dataset_class_wise_cnn_bert}, we report class-wise performance of both CNN and BERT models for the humanitarian task. BERT performs better than or on par with CNN across all classes. More importantly, BERT performs substantially better than CNN in the case of minority classes as highlighted in the table.
We further investigate the classification results of the CNN models for the minority class labels. 
We observe that the class ``response efforts'' is mostly confused with ``donation and volunteering'' and ``not humanitarian''. For example, the following tweet with ``response efforts'' label, \textit{``I am supporting Rebuild Sankhu @crowdfunderuk \#crowdfunder http://t.co/WBsKGZHHSj''}, is classified as ``donation and volunteering''. We also observe similar phenomena in minority class labels. The class ``displaced and evacuations'' is confused with ``donation and volunteering'' and ``caution and advice''. It is interesting that the class ``missing and found people'' is confused with ``donation and volunteering'' and ``not humanitarian''. The following ``missing and found people'' tweet, \textit{``RT @Fahdhusain: 11 kids recovered alive from under earthquake rubble in Awaran. Shukar Allah!!''}, is classified as ``donation and volunteering''.

\begin{table}[]
\centering
\setlength{\tabcolsep}{3pt}
\scalebox{0.74}{
\begin{tabular}{@{}lrrrrrr@{}}
\toprule
 & \multicolumn{3}{c}{\textbf{CNN}} & \multicolumn{3}{c}{\textbf{BERT}} \\ \cmidrule(lr){2-4} \cmidrule(lr){5-7}
\multicolumn{1}{c}{\textbf{Class}} & \multicolumn{1}{c}{\textbf{P}} & \multicolumn{1}{c}{\textbf{R}} & \multicolumn{1}{c}{\textbf{F1}} & \multicolumn{1}{c}{\textbf{P}} & \multicolumn{1}{c}{\textbf{R}} & \multicolumn{1}{c}{\textbf{F1}} \\ \midrule
Affected individual & 0.760 & 0.720 & 0.740 & 0.752 & 0.808 & 0.779 \\
Caution and advice & 0.630 & 0.630 & 0.630 & 0.675 & 0.707 & 0.691 \\
\rowcolor[HTML]{DAE8FC} 
Displaced and evacuations & 0.490 & 0.180 & 0.260 & 0.491 & 0.535 & 0.512 \\
Donation and volunteering & 0.700 & 0.790 & 0.740 & 0.764 & 0.807 & 0.785 \\
Infrastructure and utilities damage & 0.650 & 0.660 & 0.660 & 0.696 & 0.717 & 0.706 \\
Injured or dead people & 0.760 & 0.780 & 0.770 & 0.812 & 0.845 & 0.828 \\
\rowcolor[HTML]{DAE8FC} 
Missing and found people & 0.470 & 0.170 & 0.240 & 0.457 & 0.466 & 0.462 \\
Not humanitarian & 0.900 & 0.930 & 0.920 & 0.934 & 0.920 & 0.927 \\
Requests or needs & 0.850 & 0.840 & 0.850 & 0.909 & 0.901 & 0.905 \\
\rowcolor[HTML]{DAE8FC} 
Response efforts & 0.330 & 0.070 & 0.120 & 0.349 & 0.308 & 0.327 \\
Sympathy and support & 0.760 & 0.640 & 0.690 & 0.751 & 0.725 & 0.738 \\ \bottomrule
\end{tabular}
}
% \vspace{-0.3em}
\caption{Class-wise classification results of \textit{\textbf{humanitarian task}} on the \textit{\textbf{consolidated dataset}} using CNN and BERT.}
\label{tab:combined_dataset_class_wise_cnn_bert}
% \vspace{-0.8em}
\end{table}

\subsection{Event-aware}
\label{ssec:res_event_aware}
% \vspace{-0.2cm}
In Table~\ref{table:results_event_aware}, we report the results of the event-aware training using both CNN and BERT. The event-aware training improves the classification performance by 1.3 points (F1) using CNN for the humanitarian task compared to the results without using event information (see Table  \ref{tab:individual_combined_datasets_cnn}). However, no improvement has been observed for the informativeness task. The training using event information enables the system to use data of all disasters while preserving the disaster-specific distribution. Event-aware training is also effective in the advent of a new disaster event. Based on the type of a new disaster, one may use appropriate tags to optimize the classification performance. The event-aware training can be extended to use more than one tags. For example, in addition to preserving the event information, one can also append a tag for the disaster region. In this way, one can optimize the model for more fine-grained domain information. The event-aware training with BERT does not provide better results in any of the tasks, which requires further investigation and we leave it as a future study.

\begin{table}[]
\centering
\setlength{\tabcolsep}{3pt}
\scalebox{0.74}{
\begin{tabular}{@{}lrrrrrlll@{}}
\toprule
\multicolumn{1}{c}{\textbf{}} & \multicolumn{4}{c}{\textbf{Informativeness}} & \multicolumn{4}{c}{\textbf{Humanitarian}} \\ \cmidrule(lr){2-5} \cmidrule(lr){6-9}
\multicolumn{1}{c}{\textbf{Model}} &
  \multicolumn{1}{c}{\textbf{Acc}} &
  \multicolumn{1}{c}{\textbf{P}} &
  \multicolumn{1}{c}{\textbf{R}} &
  \multicolumn{1}{c}{\textbf{F1}} &
  \multicolumn{1}{c}{\textbf{Acc}} &
  \multicolumn{1}{c}{\textbf{P}} &
  \multicolumn{1}{c}{\textbf{R}} &
  \multicolumn{1}{c}{\textbf{F1}} \\\midrule
CNN                  & 0.868   & 0.868   & 0.870  & 0.867  & 0.847  & 0.841  & 0.850  & 0.842 \\
fastText             & 0.824   & 0.824   & 0.824  & 0.824  & 0.795  & 0.794  & 0.795  & 0.794 \\
BERT                 & 0.872   & 0.872   & 0.872  & \textbf{0.872}  & 0.865  & 0.866  & 0.865  & 0.865 \\
DistilBERT           & 0.875   & 0.874   & 0.875  & \textbf{0.874}  & 0.864  & 0.863  & 0.864  & 0.863 \\
RoBERTa              & 0.879   & 0.879   & 0.879  & \textbf{0.878}  & 0.870  & 0.871  & 0.870  & \textbf{0.870} \\ \bottomrule
\end{tabular}
}
% \vspace{-0.3em}
\caption{Results of event-aware experiments using the \textit{\textbf{consolidated dataset}}.}
\label{table:results_event_aware}
% \vspace{-1.0em}
\end{table}

\begin{table}[]
\centering
\setlength{\tabcolsep}{3pt}
\scalebox{0.74}{
\begin{tabular}{@{}lrrrrrrrr@{}}
\toprule
\multicolumn{1}{c}{\textbf{}} & \multicolumn{4}{c}{\textbf{Informativeness}} & \multicolumn{4}{c}{\textbf{Humanitarian}} \\ \midrule
\multicolumn{1}{c}{\textbf{Model}} & \multicolumn{1}{c}{\textbf{Acc}} & \multicolumn{1}{c}{\textbf{P}} & \multicolumn{1}{c}{\textbf{R}} & \multicolumn{1}{c}{\textbf{F1}} & \multicolumn{1}{c}{\textbf{Acc}} & \multicolumn{1}{c}{\textbf{P}} & \multicolumn{1}{c}{\textbf{R}} & \multicolumn{1}{c}{\textbf{F1}} \\ \midrule
\multicolumn{9}{c}{\textbf{Monolingual model}} \\ \midrule
CNN & 0.828 & 0.827 & 0.828 & 0.828 & 0.647 & 0.650 & 0.647 & 0.648 \\
fastText & 0.821 & 0.820 & 0.821 & 0.820 & 0.663 & 0.662 & 0.663 & 0.662 \\
BERT & 0.873 & 0.872 & 0.873 & 0.872 & 0.772 & 0.771 & 0.772 & 0.771 \\
DistilBERT & 0.872 & 0.871 & 0.872 & 0.871 & 0.771 & 0.770 & 0.771 & 0.770 \\
RoBERTa & 0.880 & 0.879 & 0.880 & 0.879 & 0.784 & 0.785 & 0.784 & 0.784 \\ \midrule
\multicolumn{9}{c}{\textbf{Multilingual model}} \\ \midrule
BERT-m & 0.879 & 0.879 & 0.879 & 0.879 & 0.781 & 0.783 & 0.781 & \textbf{0.781} \\
DistilBERT-m & 0.873 & 0.872 & 0.873 & 0.872 & 0.772 & 0.771 & 0.772 & 0.771 \\
XLM-RoBERTa & 0.879 & 0.879 & 0.879 & 0.879 & 0.788 & 0.789 & 0.788 & 0.788 \\ \bottomrule
\end{tabular}
}
% \vspace{-0.5em}
\caption{Results of consolidated (\textit{\textbf{multilingual}}) datasets (class label distributions are reported in Table \ref{table:data_info_source_dist} and \ref{table:data_hum_source_dist}) for both tasks and different mono and multilingual models. BERT-m: bert-base-multilingual-uncased, DistilBERT-m: distilbert-base-multilingual-cased}
\label{tab:classification_results_all_data}
% \vspace{-1.0em}
\end{table}

\subsection{Discussions}
\label{ssec:discussions}
% \vspace{-0.4em}
Social media data is noisy and it often poses a challenge for labeling and training classifiers. Our analysis on publicly available datasets reveals that one should follow a number of steps before preparing and labeling any social media dataset, not just the dataset for crisis computing. Such steps include {\em(i)} tokenization to help in the subsequent phase, {\em(ii)} remove exact- and near-duplicates, {\em(iii)} check for existing data where the same tweet might be annotated for the same task, and then {\em(iv)} labeling. For designing the classifier, we postulate checking the overlap between training and test splits to avoid any misleading performance. 

The classification performance that we report is considered as benchmark results, which can be used to compare in any future study. The current state-of-art for informativeness and humanitarian tasks can be found in \cite{burel2017semantics,Alam2019,alam2021humaid}. The F-measure for informativeness and humanitarian tasks are reported as 0.838 and 0.613, respectively, on the CrisisLex26 dataset in~\cite{burel2017semantics}. Whereas in \cite{Alam2019}, the reported F-measure for informativeness and humanitarian tasks are 0.93 and 0.78, respectively. In \cite{alam2021humaid}, the best reported F-measure for humanitarian task is 0.781. It is important to emphasize the fact that the results reported in this study are reliable as they are obtained on a dataset that has been cleaned from duplicate content, which might have led to misleading performance results otherwise. 

Our initial consolidated datasets (i.e., Table \ref{table:data_info_source_dist} and \ref{table:data_hum_source_dist}) contains multilingual content with more class labels and different types of content (e.g., disease-related), therefore, an interesting future research could be to try different pre-trained multilingual models to classify tweets in different languages. We have run a set of preliminary experiments using our initial consolidated datasets, and using monolingual model such as CNN, fastText, BERT, DistilBERT, RoBERTa, and multilingual versions of the mentioned transformer models.
% BERT, DistilBERT, RoBERTa (XLM-RoBERTa). 
The results are reported in Table \ref{tab:classification_results_all_data}. We observe that performance dropped significantly for the humanitarian task compared to English-only dataset. For example, $\sim$8\% drop using BERT model. Note that test set for English tweets (Table \ref{tab:data_split_info_hum_tasks_combined_filtered}) is a subset of this set of tweets. %Detail results is reported in the supplemental document (Section \ref{ssec:results_all_dataset} and Table \ref{tab:classification_results_all_data}). 
% From the results, one might argue that multilingual BERT might have been useful. 
From the results of multilingual versions of BERT (BERT-m),  we see that there is an increase in performance %DistilBERT (DistilBERT-m), RoBERTa (XLM-RoBERTa) $\sim$1\% gain 
compared to other mono-lingual models, however, the results are still far below. Such a finding shows an interesting avenue for further research. Another future research direction would be to use multilingual models for the zero-shot classification of tweets. 

% From the event-aware experiments, we see that it helps to improve the classification performance, which could also be a future research avenue. 

The competitive performance of transformer based models encourages us to try deeper models such as Google T5~\cite{raffel2019exploring}. For the transformer based model, it is important to invest the effort to try different regularization methods to obtain better results, which we foresee as a future study. 

Our released dataset and benchmark results will help the research community to develop better models and compare results. The inclusion of language tags can help to conduct multilingual experiments in future research. The resulting dataset covers a time-span starting from 2010 to 2017, which can be used to study temporal aspects in crisis scenarios.

%  \vspace{-0.1cm}
\section{Conclusions}
\label{sec:conclutions}
% \vspace{-0.2cm}
The information available on social media has been widely used by humanitarian organizations at times of a disaster. Many techniques and systems have been developed to process social media data. However, the research community lacks a standard dataset and benchmarks to compare the performance of their systems. We tried to bridge this gap by consolidating existing datasets, filtering exact- and near-duplicates, and providing benchmarks based on state-of-the-art CNN, FastText, and transformer-based models. Our experimental results and data splits can be useful for future research in order to conduct multilingual studies, developing new models and cross-domain experiments.
%%The dataset is provided at \href{https://ANONYMIZED}{ANONYMIZED}.

% \section*{Acknowledgments}

% The acknowledgments should go immediately before the references. Do not number the acknowledgments section.
% Do not include this section when submitting your paper for review.

\section*{Broader Impact}
The developed consolidated labeled dataset is curated from different publicly available sources.
%, which were collected from Twitter using Twitter streaming API by following its terms of service. 
The consolidated labeled dataset can be used to develop models for humanitarian response tasks and can be useful to fast responders. We release the dataset by maintaining the license of existing resources.
%and maintaining Twitter terms and services. 
% We try to maintain FAIR principles,\footnote{\url{http://www.force11.org/group/fairgroup/fairprinciples}} and a datasheet \cite{gebru2018datasheets} will be available on the data package website.
%at \textit{\url{https://crisisnlp.qcri.org/crisis_datasets_benchmarks.html}}.

{
\small
\bibliography{bib/main_bib}
}

\clearpage
\appendix
\section*{Appendix}
% \label{sec:appendix}
\label{sec:supplemental}
\section{Events and class label mapping}
\label{ssec:mapped-class-label}

In Table \ref{table:data_sources_crisislex_nlp_others_group}, we report the events associated with the respective datasets such as ISCRAM2013, SWDM2013 CrisisLex and CrisisNLP. The time-period is from 2011 to 2015, which is a good representative of temporal aspects. In Table \ref{table:class_label_mapping_crisislex_nlp_others_group}, we report class label mapping for ISCRAM2013, SWDM2013, CrisisLex and CrisisNLP datasets. The first column in the table \ref{table:class_label_mapping_crisislex_nlp_others_group} shows the mapped class for both informative and humanitarian tasks. Note that all humanitarian class labels also mapped to informative, and not humanitarian labels mapped to not-informative in the data preparation step. 
In Table \ref{table:data_sources_with_filtering_drd}, we report the class label mapping for informativeness and humanitarian tasks for DRD dataset. The DSM dataset only contains tweets labeled as relevant vs not-relevant, which we mapped for informativeness task as shown in Table \ref{table:data_sources_with_filtering_dsm}. The CrisisMMD dataset has been annotated for informativeness and humanitarian task, therefore, very minor label mapping was needed as shown in Table in \ref{table:data_sources_with_filtering_crisismmd}. The AIDR data has been labeled by domain experts using AIDR system and has been labeled during different events. The label names we mapped for informativeness and humanitarian tasks are shown in Table \ref{table:data_sources_with_filtering_aidr}. 

\begin{table*}[]
\centering
\scalebox{0.70}{
\begin{tabular}{@{}lll@{}}
\toprule
\multicolumn{1}{@{}l}{\textbf{Dataset}} & \multicolumn{1}{l}{\textbf{Year}} & \multicolumn{1}{l}{\textbf{Event name}} \\ \midrule
\multicolumn{3}{c}{\textbf{ISCRAM2013}} \\\midrule
ISCRAM2013 & 2011 & Joplin \\\midrule
\multicolumn{3}{c}{\textbf{SWDM2013}} \\\midrule
SWDM2013 & 2012 & Sandy \\\midrule
\multicolumn{3}{c}{\textbf{CrisisLex}} \\\midrule
CrisisLexT6 & 2012 & US\_Sandy Hurricane \\
CrisisLexT6 & 2013 & Alberta Floods \\
CrisisLexT6 & 2013 & Boston Bombings \\
CrisisLexT6 & 2013 & Oklahoma Tornado \\
CrisisLexT6 & 2013 & Queensland Floods \\
CrisisLexT6 & 2013 & West Texas Explosion \\
CrisisLexT26 & 2012 & Costa-Rica Earthquake \\
CrisisLexT26 & 2012 & Italy Earthquakes \\
CrisisLexT26 & 2012 & Philipinnes Floods \\
CrisisLexT26 & 2012 & Philippines Typhoon Pablo \\
CrisisLexT26 & 2012 & Venezuela Refinery Explosion \\
CrisisLexT26 & 2012 & Guatemala Earthquake \\
CrisisLexT26 & 2012 & Colorado Wildfires \\
CrisisLexT26 & 2013 & Alberta Floods \\
CrisisLexT26 & 2013 & Australia Bushfire \\
CrisisLexT26 & 2013 & Bangladesh Savar building collapse \\
CrisisLexT26 & 2013 & Bohol Earthquake \\
CrisisLexT26 & 2013 & Boston Bombings \\
CrisisLexT26 & 2013 & Brazil Nightclub Fire \\
CrisisLexT26 & 2013 & Canada Lac Megantic Train Crash \\
CrisisLexT26 & 2013 & Colorado Floods \\
CrisisLexT26 & 2013 & Glasgow Helicopter Crash \\
CrisisLexT26 & 2013 & Italy Sardinia Floods \\
CrisisLexT26 & 2013 & LA Airport Shootings \\
CrisisLexT26 & 2013 & Manila Floods \\
CrisisLexT26 & 2013 & NY Train Crash \\
CrisisLexT26 & 2013 & Phillipines Typhoon Yolanda \\
CrisisLexT26 & 2013 & Queensland Floods \\
CrisisLexT26 & 2013 & Singapore haze \\
CrisisLexT26 & 2013 & West-Texas explosion \\\midrule
\multicolumn{3}{c}{\textbf{CrisisNLP}} \\\midrule
CrisisNLP-CF & 2013 & Pakistan Earthquake \\
CrisisNLP-CF & 2014 & California Earthquake \\
CrisisNLP-CF & 2014 & Chile Earthquake \\
CrisisNLP-CF & 2014 & India Floods \\
CrisisNLP-CF & 2014 & Mexico Hurricane Odile \\
CrisisNLP-CF & 2014 & Middle-East Respiratory Syndrome \\
CrisisNLP-CF & 2014 & Pakistan Floods \\
CrisisNLP-CF & 2014 & Philippines Typhoon Hagupit \\
CrisisNLP-CF & 2014 & Worldwide Ebola \\
CrisisNLP-CF & 2015 & Nepal Earthquake \\
CrisisNLP-CF & 2015 & Vanuatu Cyclone Pam \\
CrisisNLP-volunteers & 2014-2015 & Worldwide Landslides \\
CrisisNLP-volunteers & 2014 & California Earthquake \\
CrisisNLP-volunteers & 2014 & Chile Earthquake \\
CrisisNLP-volunteers & 2014 & Iceland Volcano \\
CrisisNLP-volunteers & 2014 & Malaysia Airline MH370 \\
CrisisNLP-volunteers & 2014 & Mexico Hurricane Odile \\
CrisisNLP-volunteers & 2014 & Middle-East Respiratory Syndrome \\
CrisisNLP-volunteers & 2014 & Philippines Typhoon Hagupit \\
CrisisNLP-volunteers & 2015 & Nepal Earthquake \\
CrisisNLP-volunteers & 2015 & Vanuatu Cyclone Pam \\ \bottomrule
\end{tabular}
}
\caption{Events in CrisisLex, CrisisNLP, ISCRAM2013 and SWDM2013 datasets.}
\vspace{-0.3cm}
\label{table:data_sources_crisislex_nlp_others_group}
\end{table*}

\begin{table*}[t]
\centering
% \resizebox{\columnwidth}{!}{
\scalebox{0.55}{
\begin{tabular}{@{}lllll@{}}
\toprule
\multicolumn{1}{@{}l}{\textbf{Mapped class}} & \multicolumn{1}{l}{\textbf{Original class}} & \multicolumn{1}{l}{\textbf{Source}} & \multicolumn{1}{l}{\textbf{Annotation Description}} \\ \midrule
Affected individual & Affected individuals & CrisisLexT26 & \parbox{15cm}{Deaths, injuries, missing, found, or displaced people, and/or personal updates.}  \\
\xmark & Animal management & CrisisNLP-volunteers & \parbox{15cm}{Pets and animals, living, missing, displaced, or injured/dead} \\
Caution and advice & Caution and advice & CrisisLexT26 & \parbox{15cm}{If a message conveys/reports information about some warning or a piece of advice about a possible hazard of an incident.}  \\
Disease related & Disease signs or symptoms & CrisisNLP-CF & \parbox{15cm}{Reports of symptoms such as fever, cough, diarrhea, and shortness of breath or questions related to these symptoms.}  \\
Disease related & Disease transmission & CrisisNLP-CF & \parbox{15cm}{Reports of disease transmission or questions related to disease transmission} \\
Disease related & Disease Treatment & CrisisNLP-CF & \parbox{15cm}{Questions or suggestions regarding the treatments of the disease.}  \\
Disease related & Disease Prevention & CrisisNLP-CF & \parbox{15cm}{Questions or suggestions related to the prevention of disease or mention of a new prevention strategy.}  \\
Disease related & Disease Affected people & CrisisNLP-CF & \parbox{15cm}{Reports of affected people due to the disease} \\
Displaced and evacuations & Displaced people & CrisisNLP-volunteers & \parbox{15cm}{People who have relocated due to the crisis, even for a short time (includes evacuations)}  \\
Displaced and evacuations & Displaced people and evacuations & CrisisNLP-CF & \parbox{15cm}{People who have relocated due to the crisis, even for a short time (includes evacuations)}  \\
Donation and volunteering & \parbox{5cm}{Donation needs or offers or volunteering services} & CrisisNLP-CF & \parbox{15cm}{Reports of urgent needs or donations of shelter and/or supplies such as food, water, clothing, money, medical supplies or blood; and volunteering services} \\
Donation and volunteering & Donations and volunteering & CrisisLexT26 & \parbox{15cm}{Needs, requests, or offers of money, blood, shelter, supplies, and/or services by volunteers or professionals.} \\
Donation and volunteering & Donations of money & CrisisNLP-volunteers & Donations of money \\
Donation and volunteering & \parbox{5cm}{Donations of money goods or services} & SWDM2013/ISCRAM2013 & \parbox{15cm}{If a message speaks about money raised, donation offers, goods/services offered or asked by the victims of an incident.} \\
Donation and volunteering & \parbox{5cm}{Donations of supplies and or volunteer work} & CrisisNLP-volunteers & \parbox{15cm}{Donations of supplies and/or volunteer work} \\
Donation and volunteering & Money & CrisisNLP-volunteers & Money requested, donated or spent \\
Donation and volunteering & Shelter and supplies & CrisisNLP-volunteers & \parbox{15cm}{Needs or donations of shelter and/or supplies such as food, water, clothing, medical supplies or blood}  \\
Donation and volunteering & Volunteer or professional services & CrisisNLP-volunteers & Services needed or offered by volunteers or professionals  \\
Informative & Informative & CrisisNLP-CF & 2014 Iceland Volcano en, 2014 Malaysia Airline MH370 en  \\
Informative & Informative direct & SWDM2013/ISCRAM2013 & \parbox{15cm}{If the message is of interest to other people beyond the author's immediate circle, and seems to be written by a person who is a direct eyewitness of what is taking place.} \\
Informative & Informative direct or indirect & SWDM2013/ISCRAM2013 & \parbox{15cm}{If the message is of interest to other people beyond the author's immediate circle, but there is not enough information to tell if it is a direct report or a repetition of something from another source.}  \\
Informative & Informative indirect & SWDM2013/ISCRAM2013 & \parbox{15cm}{If the message is of interest to other people beyond the author's immediate circle, and seems to be seen/heard by the person on the radio, TV, newspaper, or other source. The message must specify the source.}  \\
Informative & related and informative & CrisisLexT26 & \parbox{15cm}{Related to the crisis and informative: if it contains useful information that helps understand the crisis situation.}  \\
Infrastructure and utilities damage & Infrastructure damage & CrisisNLP-volunteers & \parbox{15cm}{Houses, buildings, roads damaged or utilities such as water, electricity, interrupted}  \\
% Infrastructure and utilities damage & Damage & Other & \parbox{15cm}{Houses, buildings, roads damaged or utilities such as water, electricity, interrupted.}  \\
Infrastructure and utilities damage & Infrastructure and utilities & CrisisNLP-volunteers & \parbox{15cm}{Buildings or roads damaged or operational; utilities/services interrupted or restored}  \\
Infrastructure and utilities damage & Infrastructure & CrisisNLP-volunteers & Infrastructure  \\
Infrastructure and utilities damage & Infrastructure and utilities damage & CrisisNLP-CF & \parbox{15cm}{Reports of damaged buildings, roads, bridges, or utilities/services interrupted or restored.}  \\
Injured or dead people & Injured or dead people & CrisisNLP-CF & \parbox{15cm}{Reports of casualties and/or injured people due to the crisis.}  \\
Injured or dead people & Injured and dead & CrisisNLP-volunteers & Injured and dead  \\
Injured or dead people & Deaths reports & CrisisNLP-CF & Injured and dead  \\
Injured or dead people & Casualties and damage & SWDM2013/ISCRAM2013 & \parbox{15cm}{If a message reports the information about casualties or damage done by an incident.}  \\
Missing and found people & Missing trapped or found people & CrisisNLP-volunteers & \parbox{15cm}{Missing, trapped, or found people---Questions and/or reports about missing or found people.}  \\
Missing and found people & People Missing or found & CrisisNLP-volunteers & People missing or found.  \\
Missing and found people & People Missing found or seen & CrisisNLP-volunteers & \parbox{15cm}{If a message reports about the missing or found person effected by an incident or seen a celebrity visit on ground zero.}  \\
Not humanitarian & Not applicable & CrisisLexT26 & Not applicable  \\
Not humanitarian & Not related to crisis & CrisisNLP-volunteers & Not related to this crisis  \\
Not humanitarian & Not informative & \parbox{3cm}{CrisisNLP-volunteers, CrisisLexT26} & \parbox{15cm}{1. Refers to the crisis, but does not contain useful information that helps you understand the situation; 2. Not related to the Typhoon, or not relevant for emergency/humanitarian response; 3. Related to the crisis, but not informative: if it refers to the crisis, but does not contain useful information that helps understand the situation.}  \\
\xmark & Not labeled & CrisisLexT26 & Not labeled  \\
Not humanitarian & Not related or irrelevant & \parbox{3cm}{CrisisNLP-CF, CrisisNLP-volunteers} & 1. Not related or irrelevant; 2. Unrelated to the situation or irrelevant  \\
Not humanitarian & Not related to the crisis & CrisisNLP-volunteers & Not related to crisis  \\
% Not humanitarian & No damage & Other & No damage  \\
Not humanitarian & Not relevant & \parbox{3cm}{CrisisLexT26} & Not relevant  \\
Not humanitarian & Off-topic & CrisisLexT6; & Off-topic  \\
Not humanitarian & Other & CrisisNLP-volunteers & if the message is not in English, or if it cannot be classified.  \\
Not humanitarian & Not related & CrisisLexT26 & Not related  \\
% Not related or irrelevant & No damage & Cresci-SWDM15 & no damage \\
Not humanitarian & Not physical landslide & CrisisNLP-volunteers & The item does not refer to a physical landslide \\
Not humanitarian & Terrorism not related & CrisisNLP-volunteers & If the tweet is not about terrorism related to the flight MH370  \\
Other relevant information & Other relevant information & CrisisNLP-volunteers & \parbox{15cm}{1. Other useful information that helps understand the situation; 2. Informative for emergency/humanitarian response, but in none of the above categories, including weather/evacuations/etc.}  \\
Other relevant information & Other relevant & CrisisNLP-volunteers & \parbox{15cm}{1. Other useful information that helps understand the situation; 2. Informative for emergency/humanitarian response, but in none of the above categories, including weather/evacuations/etc.}  \\
Other relevant information & Other useful information & CrisisLexT26 & \parbox{15cm}{1. Other useful information not covered by any of the following categories: affected individuals, infrastructure and utilities, donations and volunteering, caution and advice, sympathy and emotional support.}  \\
Other relevant information & Related but not informative & CrisisLexT26 & \parbox{15cm}{Related to the crisis, but not informative: if it refers to the crisis, but does not contain useful information that helps understand the situation.}  \\
Other relevant information & Relevant & \parbox{3cm}{CrisisLexT26; CrisisNLP} & Relevant  \\

Personal update & Personal & CrisisNLP-volunteers & \parbox{15cm}{If the tweet conveys some sort of personal opinion, which is not of interest of a general audience.}  \\
Personal update & Personal only & CrisisNLP-volunteers & \parbox{15cm}{1. Personal and only useful to a small circle of family/friends of the author.; 2. If a message is only of interest to its author and her immediate circle of family/friends and does not convey any useful information to other people who do not know the author.}  \\
Personal update & Personal updates & CrisisNLP-volunteers & \parbox{15cm}{1. Status updates about individuals or loved ones.}  \\
Physical landslide & Physical landslide & CrisisNLP-volunteers & \parbox{15cm}{The item is related to a physical landslide} \\
Requests or needs & Needs of those affected & CrisisNLP-volunteers & Needs of those affected  \\
Requests or needs & Requests for help needs & CrisisNLP-volunteers & \parbox{15cm}{Something (e.g. food, water, shelter) or someone (e.g. volunteers, doctors) is needed.}  \\
Requests or needs & Urgent needs & CrisisNLP-volunteers & \parbox{15cm}{Something (e.g. food, water, shelter) or someone (e.g. volunteers, doctors) is needed.}  \\
Response efforts & Humanitarian aid provided & CrisisNLP-volunteers & \parbox{15cm}{Affected populations receiving food, water, shelter, medication, etc. from humanitarian/emergency response organizations.}   \\
Response efforts & Response efforts & CrisisNLP-volunteers & \parbox{15cm}{All info about responders. Affected populations receiving food, water, shelter, medication, etc. from humanitarian/emergency response organizations.} \\
Sympathy and support & Sympathy and emotional support & CrisisNLP-volunteers & Sympathy and emotional support \\
Sympathy and support & Sympathy and support & CrisisLexT26 & \parbox{15cm}{1. Thoughts, prayers, gratitude, sadness, etc.}  \\
Sympathy and support & Personal updates sympathy support & CrisisNLP-volunteers & \parbox{15cm}{Personal updates, sympathy, support.}  \\
Sympathy and support & Praying & CrisisNLP-volunteers & \parbox{15cm}{If author of the tweet prays for flight MH370 passengers.}  \\
Terrorism related information & Terrorism related & CrisisNLP-volunteers & \parbox{15cm}{If the tweet reports possible terrorism act involved.}  \\ \bottomrule
\end{tabular}
}
\caption{Class label mapping and grouping for CrisisLex, CrisisNLP, ISCRAM2013, and SWDM2013 datasets. The symbol (\xmark) indicates we do not map the tweets with that label for this study.}
\vspace{-0.3cm}
\label{table:class_label_mapping_crisislex_nlp_others_group}
\end{table*}

\begin{table*}[!tbh]
\centering
% \resizebox{\columnwidth}{!}{
\scalebox{1.0}{
\begin{tabular}{@{}lll@{}}
\toprule
\multicolumn{1}{@{}l}{\textbf{Original class}} & \multicolumn{2}{l}{\textbf{Class label mapping}} \\ \midrule
 & \multicolumn{1}{l}{\textbf{Informative}} & \multicolumn{1}{l}{\textbf{Humanitarian}} \\\cmidrule{2-3}
Related & Informative & \xmark \\
Aid related & Informative & Requests or needs \\
Request & Informative & Requests or needs \\
Offer & Informative & Donation and volunteering \\
Medical help & Informative & Requests or needs \\
Medical products & Informative & requests or needs \\
Search and rescue & Informative & displaced and evacuations \\
Security & \xmark & \xmark \\
Military & \xmark & \xmark \\
Water & Informative & Requests or needs \\
Food & Informative & Requests or needs \\
Shelter & Informative & Requests or needs \\
Clothing & Informative & Requests or needs \\
Money & Informative & Requests or needs \\
Missing people & Informative & Missing and found people \\
Refugees & Informative & Requests or needs \\
Death & Informative & Injured or dead people \\
Other aid & Informative & Requests or needs \\
Infrastructure related & Informative & Infrastructure and Utilities damage \\
Transport & Informative & Infrastructure and utilities damage \\
Buildings & Informative & Infrastructure and utilities damage \\
Electricity & Informative & Infrastructure and utilities damage \\
Hospitals & Informative & Infrastructure and utilities damage \\
Shops & Informative & Infrastructure and utilities damage \\
Aid centers & Informative & Infrastructure and utilities damage \\
Other infrastructure & Informative & Infrastructure and Utilities damage \\ \bottomrule
\end{tabular}
}
\caption{Class label mapping for Disaster Response Data (DRD). The symbol (\xmark) indicates we do not map the tweets with that label for this study.}
\label{table:data_sources_with_filtering_drd}
\end{table*}

\begin{table*}[h]
\centering
\scalebox{1.00}{
\begin{tabular}{@{}ll@{}}
\toprule
% \multicolumn{2}{c}{\textbf{Class label mapping}} \\\midrule
\multicolumn{1}{@{}l}{\textbf{Original class}} & \multicolumn{1}{l}{\textbf{Mapped class}} \\\midrule
Relevant & Informative \\
Not Relevant & Not informative \\\bottomrule
\end{tabular}
}
\caption{Class label mapping for Disasters on Social Media (DSM) dataset.}
\label{table:data_sources_with_filtering_dsm}
\end{table*}

\begin{table*}[h]
\centering
\scalebox{1.0}{
\begin{tabular}{@{}lll@{}}
\toprule
% \multicolumn{3}{c}{\textbf{Class label mapping}} \\ \midrule
% \multicolumn{1}{c}{\textbf{Original class}}& \multicolumn{1}{c}{\textbf{Informative}} & \multicolumn{1}{c}{\textbf{Humanitarian}} \\ \midrule
\multicolumn{1}{@{}l}{\textbf{Original class}} & \multicolumn{2}{l}{\textbf{Class label mapping}} \\ \midrule
 & \multicolumn{1}{l}{\textbf{Informative}} & \multicolumn{1}{l}{\textbf{Humanitarian}} \\\cmidrule{2-3}
Affected individuals & Informative & Affected individual \\
Infrastructure and utility damage & Informative & Infrastructure and utilities damage \\
Injured or dead people & Informative & Injured or dead people \\
Missing or found people & Informative & Missing and found people \\
Not relevant or cant judge & Not informative & Not humanitarian \\
Other relevant information & Informative & Other relevant information \\
Rescue volunteering or donation effort & Informative & Donation and volunteering \\
Vehicle damage & Informative & Infrastructure and utilities damage \\ \bottomrule
\end{tabular}
}
\caption{Class label mapping for CrisisMMD.}
\label{table:data_sources_with_filtering_crisismmd}
\end{table*}

\begin{table*}[!tbh]
\centering
\scalebox{0.90}{
\begin{tabular}{@{}lll@{}}
\toprule
\multicolumn{1}{@{}l}{\textbf{Original class}} & \multicolumn{2}{l}{\textbf{Class label mapping}} \\ \midrule
 & \multicolumn{1}{l}{\textbf{Informative}} & \multicolumn{1}{l}{\textbf{Humanitarian}} \\\cmidrule{2-3}
Blocked roads & Informative & Infrastructure and utilities damage \\
Blood or other medical supplies needed & Informative & Requests or needs \\
Building damaged & Informative & Infrastructure and utilities damage \\
Camp shelter & Informative & Requests or needs \\
Casualties and damage & Informative & Infrastructure and utilities damage \\
Caution and advice & Informative & Caution and advice \\
Clothing needed & Informative & Requests or needs \\
Damage & Informative & Infrastructure and utilities damage \\
Displaced people & Informative & Displaced and evacuations \\
Donations & Informative & Donation and volunteering \\
Food and or water needed & Informative & Requests or needs \\
Food water & Informative & Requests or needs \\
Humanitarian aid provided & Informative & Response efforts \\
Informative & Informative & Informative \\
Infrastructure and utilities & Informative & Infrastructure and utilities damage \\
Infrastructure damage & Informative & Infrastructure and utilities damage \\
Injured dead & Informative & Injured or dead people \\
Injured or dead people & Informative & Injured or dead people \\
Loss of electricity & Informative & Infrastructure and utilities damage \\
Loss of internet & Informative & Infrastructure and utilities damage \\
Missing trapped or found people & Informative & Missing and found people \\
Money & Informative & Requests or needs \\
Money needed & Informative & Requests or needs \\
Needs and requests for help & Informative & Requests or needs \\
Non emergency but relevant & Informative & \xmark \\
None of the above & Not informative & Not humanitarian \\
Not informative & Not informative & Not humanitarian \\
Not related or irrelevant & Not informative & Not humanitarian \\
Not relevant & Not informative & Not humanitarian \\
Not relevent & Not informative & Not humanitarian \\
Other relevant & Informative & Other relevant information \\
Other relevant information & Informative & Other relevant information \\
Other useful for response & Informative & Other relevant information \\
Relief and response efforts & Informative & Requests or needs \\
Requests for help needs & Informative & Requests or needs \\
Response efforts & Informative & Requests or needs \\
Shelter & Informative & Requests or needs \\
Shelter and supplies & Informative & Requests or needs \\
Shelter needed & Informative & Requests or needs \\
Shelter or supplies needed & Informative & Requests or needs \\
Sympathy and emotional support & Informative & Sympathy and support \\
Urgent needs & Informative & Requests or needs \\ \bottomrule
\end{tabular}
}
\caption{Class label mapping for AIDR system.}
%and filtering some class labels
\label{table:data_sources_with_filtering_aidr}
\end{table*}

% \begin{table*}[]
% \centering
% \scalebox{0.80}{
% \begin{tabular}{@{}llll@{}}
% \toprule
% \multicolumn{1}{c}{\textbf{\#}} & \multicolumn{1}{c}{\textbf{Tweet}} & \multicolumn{1}{c}{\textbf{Tokenized}} & \multicolumn{1}{c}{\textbf{Sim}} \\ \midrule
% \multirow{2}{*}{1} & \parbox{8cm}{When i see people praying Queensland won't flood to save a flash mob, instead of people's lives and homes. S E L F I S H} & \parbox{8cm}{when i see people praying queensland wo n't flood to save a flash mob instead of people 's lives and homes s e l f i s h} & \multirow{2}{*}{0.983} \\
% % \rowcolor[HTML]{DAE8FC} 
%  & \cellcolor{gray!25}\parbox{8cm}{RT @nialls\_swag: When i see people praying Queensland won't flood to save a flash mob, instead of people's lives and homes. S E L F I S H} & \cellcolor{gray!25}\parbox{8cm}{rt when i see people praying queensland wo n't flood to save a flash mob instead of people 's lives and homes s e l f i s h} &  \\ \bottomrule
% \end{tabular}
% }
% \caption{Examples of near-duplicates with similarity scores.}
% \label{table:near-duplicates-examples}
% \end{table*}

\paragraph{Examples of Tweets with Similarity}
\label{ssec:more_sim_examples}
In Table \ref{table:near-duplicates-examples_additional}, we report example tweets with different similarity values to justify the selection of similarity threshold. We have chosen a value of $>0.75$ to filter duplicate tweets.

\begin{table*}[!tbh]
\centering
\scalebox{0.75}{
\begin{tabular}{lllll}
\toprule
\multicolumn{1}{l}{\textbf{\#}} & \multicolumn{1}{c}{\textbf{Tweet}} & \multicolumn{1}{c}{\textbf{Tokenized}} & \multicolumn{1}{c}{\textbf{Sim.}} & \multicolumn{1}{c}{\textbf{Dup.}}\\ \midrule

\multirow{2}{*}{1} & \parbox{8cm}{RT @rosemaryCNN: As flood waters recede in Qld, \#Australia, attention turns 2 relief \& recovery. Police reportedly find a 5th victim ...} & \parbox{8cm}{rt as flood waters recede in qld australia attention turns relief recovery police reportedly find a th victim} & \multirow{2}{*}{0.882} & \multirow{2}{*}{\cmark} \\
 & \cellcolor[gray]{0.8}\parbox{8cm}{As flood waters recede in Qld, \#Australia, attention turns 2 relief \& recovery. Police reportedly find a 5th victim in a car \#CNN} & \cellcolor[gray]{0.8} \parbox{8cm}{as flood waters recede in qld australia attention turns relief recovery police reportedly find a th victim in a car cnn} &  \\ \midrule
\multirow{2}{*}{2} & \parbox{8cm}{Queensland counts flood cost as New South Wales braces for river peaks - The Guardian: The GuardianQueensland co... http://t.co/PyGhSzbG} & \parbox{8cm}{queensland counts flood cost as new south wales braces for river peaks the guardian the guardianqueensland co url} & \multirow{2}{*}{0.856} & \multirow{2}{*}{\cmark}\\
 & \cellcolor[gray]{0.8}\parbox{8cm}{Queensland counts flood cost as New South Wales braces for river peaks - The Guardian http://t.co/njADhrdc \#News} & \cellcolor[gray]{0.8}\parbox{8cm}{queensland counts flood cost as new south wales braces for river peaks the guardian url news} &  \\ \midrule
\multirow{2}{*}{3} & \parbox{8cm}{He's no Anna Bligh! @abcnews LIVE: Queensland Premier Campbell Newman is giving an update on Queensland flood crisis http://t.co/pXxoxLOe} & \parbox{8cm}{he 's no anna bligh live queensland premier campbell newman is giving an update on queensland flood crisis url} & \multirow{2}{*}{0.808} & \multirow{2}{*}{\cmark}\\
 & \cellcolor[gray]{0.8}\parbox{8cm}{AUSTRALIA: RT @abcnews: LIVE: Queensland Premier Campbell Newman is giving an update on Queensland flood crisis http://t.co/Jj9S057T} & \cellcolor[gray]{0.8}\parbox{8cm}{australia rt live queensland premier campbell newman is giving an update on queensland flood crisis url} &  \\ \midrule
\multirow{2}{*}{4} & \parbox{8cm}{Australia lurches from fire to flood http://t.co/C6x8Uxnk} & \parbox{8cm}{australia lurches from fire to flood url} & \multirow{2}{*}{0.807} & \multirow{2}{*}{\cmark}\\
 & \cellcolor[gray]{0.8}\parbox{8cm}{Australia lurches from fire to flood \#climatechange \#globalwarming http://t.co/MZa6H3QC} & \cellcolor[gray]{0.8}\parbox{8cm}{australia lurches from fire to flood climatechange globalwarming url} &  \\ \midrule
\multirow{2}{*}{5} & \parbox{8cm}{Live coverage: Queensland flood crisis via @Y7News http://t.co/Knb407Fw} & \parbox{8cm}{live coverage queensland flood crisis via url} & \multirow{2}{*}{0.788} & \multirow{2}{*}{\cmark}\\
 & \cellcolor[gray]{0.8}\parbox{8cm}{Live coverage: Queensland flood crisis - Yahoo!7 http://t.co/U2hw0LWW  via @Y7News} & \cellcolor[gray]{0.8}\parbox{8cm}{live coverage queensland flood crisis yahoo url via} &  \\ \midrule
\multirow{2}{*}{6} & \parbox{8cm}{Halo tetangga. Sabar ya. RT  @AJEnglish: Flood worsens in eastern Australia http://t.co/YfokqBmG} & \parbox{8cm}{halo tetangga sabar ya rt flood worsens in eastern australia url} & \multirow{2}{*}{0.787} & \multirow{2}{*}{\cmark}\\
 & \cellcolor[gray]{0.8}\parbox{8cm}{RT @AJEnglish: Flood worsens in eastern Australia http://t.co/kuGSMCiH} & \cellcolor[gray]{0.8}\parbox{8cm}{rt flood worsens in eastern australia url} &  \\ \midrule
\multirow{2}{*}{7} & \parbox{8cm}{"@guardian: Queensland counts flood cost as New South Wales braces for river peaks http://t.co/MpQskYt1". Brisbane friends moved to refuge.} & \parbox{8cm}{queensland counts flood cost as new south wales braces for river peaks url brisbane friends moved to refuge} & \multirow{2}{*}{0.778} & \multirow{2}{*}{\cmark}\\
 & \cellcolor[gray]{0.8}\parbox{8cm}{Queensland counts flood cost as New South Wales braces for river peaks http://t.co/qb5UuYf9} & \cellcolor[gray]{0.8}\parbox{8cm}{queensland counts flood cost as new south wales braces for river peaks url} & \\ \midrule
 
 \multirow{2}{*}{8} & \parbox{8cm}{RT @FoxNews: \#BREAKING: Numerous injuries reported in large explosion at \#Texas fertilizer plant http://t.co/oH93niFiAS". Brisbane friends moved to refuge.} & \parbox{8cm}{rt breaking numerous injuries reported in large explosion at texas fertilizer plant url} & \multirow{2}{*}{0.744} & \multirow{2}{*}{\xmark}\\
 & \cellcolor[gray]{0.8}\parbox{8cm}{Numerous injuries reported in large explosion at Texas fertilizer plant: DEVELOPING: Emergency crews in Texas ... http://t.co/Th5Yzvdg5m} & \cellcolor[gray]{0.8}\parbox{8cm}{numerous injuries reported in large explosion at texas fertilizer plant developing emergency crews in texas url} & \\ \midrule
 
\multirow{2}{*}{9} & \parbox{8cm}{Obama to attend memorial service for victims of Texas explosion: The president will meet with victims of the d... http://t.co/VgGdVATn1b} & \parbox{8cm}{obama to attend memorial service for victims of texas explosion the president will meet with victims of the d url} & \multirow{2}{*}{0.732} & \multirow{2}{*}{\xmark}\\
 & \cellcolor[gray]{0.8}\parbox{8cm}{Obama to attend memorial service for victims of Texas explosion http://t.co/f6JXfzd7QZ} & \cellcolor[gray]{0.8}\parbox{8cm}{obama to attend memorial service for victims of texas explosion url} & \\ \midrule
 
\multirow{2}{*}{10} & \parbox{8cm}{RT @RobertTaylors: Shooting Reported at Los Angeles International Airport: There are reports of a shooting incident Friday mornin... http:/…} & \parbox{8cm}{rt shooting reported at los angeles international airport there are reports of a shooting incident friday mornin http …} & \multirow{2}{*}{0.705} & \multirow{2}{*}{\xmark}\\
 & \cellcolor[gray]{0.8}\parbox{8cm}{RT @BuzzFeed: There Are Reports Of A Shooting At Los Angeles International Airport http://t.co/9TgunRXajQ} & \cellcolor[gray]{0.8}\parbox{8cm}{rt there are reports of a shooting at los angeles international airport url} & \\ \midrule

 \multirow{2}{*}{11} & \parbox{8cm}{“@BuzzFeed: Watch Hurricane Sandy roll in from the top of the @nytimes building http://t.co/dl2g3sAH”} & \parbox{8cm}{watch hurricane sandy roll in from the top of the building url} & \multirow{2}{*}{0.709} & \multirow{2}{*}{\xmark}\\
 & \cellcolor[gray]{0.8}\parbox{8cm}{Hurricane Sandy view from the top of the NYTimes building http://t.co/pLiXlaHI} & \cellcolor[gray]{0.8}\parbox{8cm}{hurricane sandy view from the top of the nytimes building url} & \\
 \bottomrule 
\end{tabular}
}
\vspace{-0.5em}
\caption{Examples of near-duplicates with similarity scores selected from informativeness tweets. Duplicates are highlighted. \textit{Sim.} refers to similarity value. \textit{Dup.} refers to whether we consider them as duplicate and filtered. The symbol (\xmark) indicates a duplicate, which we dropped and the symbol (\xmark) indicates a not duplicate, which we have included in our dataset.}
\label{table:near-duplicates-examples_additional}
\end{table*}

\section{Experimental Parameters}
\label{sec:exp_parameters}
In this section, we report parameters for CNN and BERT model. In addition, we discuss the computing infrastructures that we used for the experiments. 
\subsection{CNN Parameters}

\begin{itemize}
    \item Batch size: 128
    \item Filter size of 300 
    \item Window size of 2, 3, and 4
    \item Pooling length of 2, 3, and 4
    \item Number of epochs: 3000
    \item Max seq length: 60
    \item Patience for early stopping: 200
    \item Learning rate (Adam): 1.0E-05
    \item 
\end{itemize}

\subsection{FastText Parameters}
We mostly used default parameters as can be found int our released package.
\begin{itemize}
    \item Embedding dimension: 300
    \item Minimal number of word occurrences: 3
    \item Number of epochs: 50
\end{itemize}
\subsection{Transformer Models' Parameters}
% \textbf{BERT model} (bert-base-uncased): L=12, H=768, A=12, total parameters = 110M; where \textit{L} is number of layers (i.e., Transformer blocks), \textit{H} is the hidden size, and \textit{A} is the number of self-attention heads.
Below we list the hyperparameters that we used for training across all transformer based models. All experimental scripts will be publicly 
Hyper-parameters include:
\begin{itemize}
    \item Batch size: 8
    \item Number of epochs: 10
    \item Max seq length: 128
    \item Learning rate (Adam): 2e-5
\end{itemize}

\textbf{Number of parameters:}
\begin{itemize}
    \item \textbf{BERT} (bert-base-uncased): L=12, H=768, A=12, total parameters = 110M; where \textit{L} is number of layers (i.e., Transformer blocks), \textit{H} is the hidden size, and \textit{A} is the number of self-attention heads.
    \item \textbf{DistilBERT} (distilbert-base-uncased): it is a distilled version of the BERT model consists of 6-layer, 768-hidden, 12-heads, 66M parameters.
    \item \textbf{RoBERTa} (roberta-large): it is using the BERT-large architecture consists of 24-layer, 1024-hidden, 16-heads, 355M parameters.
    %12 768 3072 12 250k 270M
\end{itemize}   

% : {\em (i)} BERT\textsubscript{bert-base-uncased} (110M), {\em (ii)} DistilBERT\textsubscript{distilbert-base-uncased} (66M), {\em (iii)} DistilBERT\textsubscript{roberta-large} (355M)

% \subsection{Computing Infrastructure}
% We used the NVIDIA Tesla P100 16 GB GPU machine consists of 32 cores and 173GB CPU memory. 

\subsection{Detail Results}
\label{ssec:results_all_dataset}
In Table \ref{tab:all_results_info} and \ref{tab:all_results_hum}, we provide detail results for different datasets (English Tweets) with different models.

\begin{table*}[!tbh]
\centering
\scalebox{0.70}{
\begin{tabular}{@{}llrrrrllrrrr@{}}
\toprule
\multicolumn{1}{c}{\textbf{Train}} & \multicolumn{1}{c}{\textbf{Test}} & \multicolumn{1}{c}{\textbf{Acc}} & \multicolumn{1}{c}{\textbf{P}} & \multicolumn{1}{c}{\textbf{R}} & \multicolumn{1}{c}{\textbf{F1}} & \multicolumn{1}{c}{\textbf{Train}} & \multicolumn{1}{c}{\textbf{Test}} & \multicolumn{1}{c}{\textbf{Acc}} & \multicolumn{1}{c}{\textbf{P}} & \multicolumn{1}{c}{\textbf{R}} & \multicolumn{1}{c}{\textbf{F1}} \\ \midrule
\multicolumn{6}{c}{\textbf{CNN}} & \multicolumn{6}{c}{\textbf{FastText}} \\ \midrule
CrisisLex & CrisisLex & 0.945 & 0.945 & 0.950 & 0.945 & CrisisLex & CrisisLex & 0.940 & 0.940 & 0.940 & 0.940 \\
CrisisLex & CrisisNLP & 0.689 & 0.688 & 0.690 & 0.689 & CrisisLex & CrisisNLP & 0.685 & 0.694 & 0.685 & 0.687 \\
CrisisLex & Consolidated & 0.801 & 0.807 & 0.800 & 0.803 & CrisisLex & Consolidated & 0.789 & 0.803 & 0.789 & 0.791 \\
CrisisNLP & CrisisNLP & 0.832 & 0.832 & 0.830 & 0.832 & CrisisNLP & CrisisNLP & 0.817 & 0.816 & 0.817 & 0.816 \\
CrisisNLP & CrisisLex & 0.712 & 0.803 & 0.701 & 0.705 & CrisisNLP & CrisisLex & 0.729 & 0.781 & 0.729 & 0.728 \\
CrisisNLP & Consolidated & 0.725 & 0.768 & 0.730 & 0.727 & CrisisNLP & Consolidated & 0.731 & 0.754 & 0.731 & 0.733 \\
Consolidated & CrisisLex & 0.943 & 0.943 & 0.940 & 0.943 & Consolidated & CrisisLex & 0.918 & 0.918 & 0.918 & 0.917 \\
Consolidated & CrisisNLP & 0.829 & 0.828 & 0.830 & 0.828 & Consolidated & CrisisNLP & 0.812 & 0.811 & 0.812 & 0.811 \\
Consolidated & Consolidated & 0.867 & 0.866 & 0.870 & 0.866 & Consolidated & Consolidated & 0.845 & 0.844 & 0.845 & 0.844 \\ \midrule
\multicolumn{6}{c}{\textbf{BERT}} & \multicolumn{6}{c}{\textbf{RoBERTa}} \\ \midrule
CrisisLex & CrisisLex & 0.949 & 0.949 & 0.949 & 0.949 & CrisisLex   (6C) & CrisisLex & 0.938 & 0.939 & 0.938 & 0.938 \\
CrisisLex & CrisisNLP & 0.704 & 0.703 & 0.704 & 0.698 & CrisisLex & CrisisNLP & 0.724 & 0.728 & 0.724 & 0.715 \\
CrisisLex & Consolidated & 0.806 & 0.808 & 0.806 & 0.806 & CrisisLex & Consolidated & 0.803 & 0.802 & 0.803 & 0.802 \\
CrisisNLP & CrisisNLP & 0.834 & 0.834 & 0.834 & 0.833 & CrisisNLP   (10C) & CrisisNLP & 0.824 & 0.823 & 0.824 & 0.823 \\
CrisisNLP & CrisisLex & 0.752 & 0.819 & 0.752 & 0.749 & CrisisNLP & CrisisLex & 0.731 & 0.811 & 0.731 & 0.726 \\
CrisisNLP & Consolidated & 0.751 & 0.782 & 0.751 & 0.753 & CrisisNLP & Consolidated & 0.742 & 0.769 & 0.742 & 0.744 \\
Consolidated & CrisisLex & 0.940 & 0.940 & 0.940 & 0.940 & Consolidated & CrisisLex & 0.946 & 0.946 & 0.946 & 0.945 \\
Consolidated & CrisisNLP & 0.825 & 0.825 & 0.825 & 0.825 & Consolidated & CrisisNLP & 0.831 & 0.832 & 0.831 & 0.830 \\
Consolidated & Consolidated & 0.872 & 0.872 & 0.872 & 0.872 & Consolidated   (11C) & Consolidated & 0.884 & 0.883 & 0.884 & 0.883 \\ \midrule
\multicolumn{6}{c}{\textbf{DistilBERT}} &  &  & \multicolumn{1}{l}{} & \multicolumn{1}{l}{} & \multicolumn{1}{l}{} & \multicolumn{1}{l}{} \\ \midrule
CrisisLex & CrisisLex & 0.949 & 0.950 & 0.949 & 0.949 &  &  & \multicolumn{1}{l}{} & \multicolumn{1}{l}{} & \multicolumn{1}{l}{} & \multicolumn{1}{l}{} \\
CrisisLex & CrisisNLP & 0.691 & 0.693 & 0.691 & 0.681 &  &  & \multicolumn{1}{l}{} & \multicolumn{1}{l}{} & \multicolumn{1}{l}{} & \multicolumn{1}{l}{} \\
CrisisLex & Consolidated & 0.808 & 0.808 & 0.808 & 0.808 &  &  & \multicolumn{1}{l}{} & \multicolumn{1}{l}{} & \multicolumn{1}{l}{} & \multicolumn{1}{l}{} \\
CrisisNLP & CrisisNLP & 0.834 & 0.834 & 0.834 & 0.834 &  &  & \multicolumn{1}{l}{} & \multicolumn{1}{l}{} & \multicolumn{1}{l}{} & \multicolumn{1}{l}{} \\
CrisisNLP & CrisisLex & 0.743 & 0.818 & 0.743 & 0.739 &  &  & \multicolumn{1}{l}{} & \multicolumn{1}{l}{} & \multicolumn{1}{l}{} & \multicolumn{1}{l}{} \\
CrisisNLP & Consolidated & 0.752 & 0.783 & 0.752 & 0.755 &  &  & \multicolumn{1}{l}{} & \multicolumn{1}{l}{} & \multicolumn{1}{l}{} & \multicolumn{1}{l}{} \\
Consolidated & CrisisLex & 0.938 & 0.938 & 0.938 & 0.938 &  &  & \multicolumn{1}{l}{} & \multicolumn{1}{l}{} & \multicolumn{1}{l}{} & \multicolumn{1}{l}{} \\
Consolidated & CrisisNLP & 0.829 & 0.828 & 0.829 & 0.828 &  &  & \multicolumn{1}{l}{} & \multicolumn{1}{l}{} & \multicolumn{1}{l}{} & \multicolumn{1}{l}{} \\
Consolidated & Consolidated & 0.871 & 0.870 & 0.871 & 0.87 &  &  & \multicolumn{1}{l}{} & \multicolumn{1}{l}{} & \multicolumn{1}{l}{} & \multicolumn{1}{l}{} \\ \bottomrule
\end{tabular}
}
\caption{Results of \textbf{\textit{Informativeness}} task using different models on \textbf{\textit{consolidated English Tweets}}.}
\label{tab:all_results_info}
\end{table*}

\begin{table*}[!tbh]
\centering
\scalebox{0.70}{
\begin{tabular}{@{}llrrrrllrrrr@{}}
\toprule
\multicolumn{1}{c}{\textbf{Train}} & \multicolumn{1}{c}{\textbf{Test}} & \multicolumn{1}{c}{\textbf{Acc}} & \multicolumn{1}{c}{\textbf{P}} & \multicolumn{1}{c}{\textbf{R}} & \multicolumn{1}{c}{\textbf{F1}} & \multicolumn{1}{c}{\textbf{Train}} & \multicolumn{1}{c}{\textbf{Test}} & \multicolumn{1}{c}{\textbf{Acc}} & \multicolumn{1}{c}{\textbf{P}} & \multicolumn{1}{c}{\textbf{R}} & \multicolumn{1}{c}{\textbf{F1}} \\\midrule
\multicolumn{6}{c}{\textbf{CNN}} & \multicolumn{6}{c}{\textbf{FastText}} \\ \midrule
CrisisLex (6C) & CrisisLex & 0.921 & 0.920 & 0.920 & 0.920 & CrisisLex   (6C) & CrisisLex & 0.914 & 0.909 & 0.914 & 0.911 \\
CrisisLex & CrisisNLP & 0.554 & 0.546 & 0.550 & 0.544 & CrisisLex & CrisisNLP & 0.578 & 0.543 & 0.578 & 0.549 \\
CrisisLex & Consolidated & 0.694 & 0.601 & 0.690 & 0.633 & CrisisLex & Consolidated & 0.670 & 0.578 & 0.670 & 0.605 \\
CrisisNLP   (10C) & CrisisNLP & 0.780 & 0.757 & 0.780 & 0.762 & CrisisNLP   (10C) & CrisisNLP & 0.774 & 0.756 & 0.774 & 0.759 \\
CrisisNLP & CrisisLex & 0.769 & 0.726 & 0.770 & 0.714 & CrisisNLP & CrisisLex & 0.773 & 0.729 & 0.773 & 0.719 \\
CrisisNLP & Consolidated & 0.666 & 0.582 & 0.670 & 0.613 & CrisisNLP & Consolidated & 0.690 & 0.616 & 0.690 & 0.627 \\
Consolidated & CrisisLex & 0.908 & 0.916 & 0.910 & 0.912 & Consolidated & CrisisLex & 0.901 & 0.906 & 0.901 & 0.903 \\
Consolidated & CrisisNLP & 0.768 & 0.753 & 0.770 & 0.753 & Consolidated & CrisisNLP & 0.769 & 0.759 & 0.769 & 0.760 \\
Consolidated   (11C) & Consolidated & 0.835 & 0.827 & 0.840 & 0.829 & Consolidated   (11C) & Consolidated & 0.830 & 0.821 & 0.830 & 0.824 \\ \midrule
\multicolumn{6}{c}{\textbf{BERT}} & \multicolumn{6}{c}{\textbf{RoBERTa}} \\ \midrule
CrisisLex (6C) & CrisisLex & 0.934 & 0.935 & 0.934 & 0.934 & CrisisLex   (6C) & CrisisLex & 0.936 & 0.938 & 0.936 & 0.937 \\
CrisisLex & CrisisNLP & 0.567 & 0.757 & 0.567 & 0.615 & CrisisLex & CrisisNLP & 0.588 & 0.753 & 0.588 & 0.632 \\
CrisisLex & Consolidated & 0.757 & 0.794 & 0.757 & 0.766 & CrisisLex & Consolidated & 0.775 & 0.810 & 0.775 & 0.784 \\
CrisisNLP   (10C) & CrisisNLP & 0.793 & 0.79 & 0.793 & 0.791 & CrisisNLP   (10C) & CrisisNLP & 0.789 & 0.790 & 0.789 & 0.789 \\
CrisisNLP & CrisisLex & 0.855 & 0.863 & 0.855 & 0.842 & CrisisNLP & CrisisLex & 0.856 & 0.876 & 0.856 & 0.850 \\
CrisisNLP & Consolidated & 0.738 & 0.717 & 0.738 & 0.709 & CrisisNLP & Consolidated & 0.746 & 0.748 & 0.746 & 0.727 \\
Consolidated & CrisisLex & 0.920 & 0.927 & 0.920 & 0.923 & Consolidated & CrisisLex & 0.927 & 0.934 & 0.927 & 0.931 \\
Consolidated & CrisisNLP & 0.785 & 0.787 & 0.785 & 0.786 & Consolidated & CrisisNLP & 0.784 & 0.785 & 0.784 & 0.784 \\
Consolidated   (11C) & Consolidated & 0.859 & 0.861 & 0.859 & 0.860 & Consolidated   (11C) & Consolidated & 0.871 & 0.873 & 0.871 & 0.872 \\ \midrule
\multicolumn{6}{c}{\textbf{DistilBERT}} & \multicolumn{1}{c}{\textbf{}} & \multicolumn{1}{c}{\textbf{}} & \multicolumn{1}{c}{\textbf{}} & \multicolumn{1}{c}{\textbf{}} & \multicolumn{1}{c}{\textbf{}} & \multicolumn{1}{c}{\textbf{}} \\ \midrule
CrisisLex (6C) & CrisisLex & 0.935 & 0.935 & 0.935 & 0.935 &  &  & \multicolumn{1}{l}{} & \multicolumn{1}{l}{} & \multicolumn{1}{l}{} & \multicolumn{1}{l}{} \\
CrisisLex & CrisisNLP & 0.579 & 0.754 & 0.579 & 0.628 &  &  & \multicolumn{1}{l}{} & \multicolumn{1}{l}{} & \multicolumn{1}{l}{} & \multicolumn{1}{l}{} \\
CrisisLex & Consolidated & 0.763 & 0.792 & 0.763 & 0.770 &  &  & \multicolumn{1}{l}{} & \multicolumn{1}{l}{} & \multicolumn{1}{l}{} & \multicolumn{1}{l}{} \\
CrisisNLP   (10C) & CrisisNLP & 0.791 & 0.786 & 0.791 & 0.788 &  &  & \multicolumn{1}{l}{} & \multicolumn{1}{l}{} & \multicolumn{1}{l}{} & \multicolumn{1}{l}{} \\
CrisisNLP & CrisisLex & 0.860 & 0.868 & 0.860 & 0.845 &  &  & \multicolumn{1}{l}{} & \multicolumn{1}{l}{} & \multicolumn{1}{l}{} & \multicolumn{1}{l}{} \\
CrisisNLP & Consolidated & 0.738 & 0.730 & 0.738 & 0.707 &  &  & \multicolumn{1}{l}{} & \multicolumn{1}{l}{} & \multicolumn{1}{l}{} & \multicolumn{1}{l}{} \\
Consolidated & CrisisLex & 0.918 & 0.924 & 0.918 & 0.921 &  &  & \multicolumn{1}{l}{} & \multicolumn{1}{l}{} & \multicolumn{1}{l}{} & \multicolumn{1}{l}{} \\
Consolidated & CrisisNLP & 0.791 & 0.786 & 0.791 & 0.787 &  &  & \multicolumn{1}{l}{} & \multicolumn{1}{l}{} & \multicolumn{1}{l}{} & \multicolumn{1}{l}{} \\
Consolidated   (11C) & Consolidated & 0.857 & 0.856 & 0.857 & 0.856 &  &  & \multicolumn{1}{l}{} & \multicolumn{1}{l}{} & \multicolumn{1}{l}{} & \multicolumn{1}{l}{} \\ \bottomrule
\end{tabular}
}
\caption{Results of \textbf{\textit{Humanitarian}} task using different models on \textit{\textbf{consolidated English Tweets}}.}
\label{tab:all_results_hum}
\end{table*}

\end{document}